\documentclass[reprint, amsmath, amssymb, aps, prb, superscriptaddress, nofootinbib]{revtex4-2}

\usepackage[dvipdfmx]{graphicx,color}
\usepackage{dcolumn}
\usepackage{bm}
\usepackage{mhchem}
\usepackage{hyperref}
\usepackage{mathrsfs}
\usepackage{physics2} 
\usepackage{braket} 
\usepackage[dvipsnames]{xcolor} 
\usepackage{txfonts}
\usepackage{cancel}

\usepackage{siunitx}
\DeclareSIUnit{\muB}{\mu_{\mathrm{B}}}
\DeclareSIUnit{\rydberg}{\mathrm{Ry}}
\DeclareSIUnit\angstrom{\mathring{\mathrm{A}}}

\usepackage{color}

\usepackage{CJKutf8}

\begin{document}

\preprint{NbMnP_NHE}

\title{Nodal-line-enhanced quantum geometric effects: anomalous and nonlinear Hall effects in the parity-mixed antiferromagnet NbMnP}
\author{Ibuki Terada}
\affiliation{
Department of Materials Science, Graduate School of Engineering, Osaka Metropolitan University, Sakai, Osaka 599-8531, Japan
}
\author{Vu Thi Ngoc Huyen}
\affiliation{
Institute for Materials Research, Tohoku University, Sendai, Miyagi 980-8577, Japan
}
\author{Yuki Yanagi}
\affiliation{
Liberal Arts and Sciences, Toyama Prefectural University, Toyama 939-0398, Japan
}
\author{Michi-To Suzuki}
\affiliation{
Department of Materials Science, Graduate School of Engineering, Osaka Metropolitan University, Sakai, Osaka 599-8531, Japan
}
\affiliation{
Center for Spintronics Research Network, Graduate School of Engineering Science,
Osaka University, Toyonaka, Osaka 560-8531, Japan
}

\date{\today}

\begin{abstract}
The anomalous Hall effect has been understood in terms of the geometric nature of Bloch bands and impurity scattering, and has been observed in a wide variety of magnetic materials such as ferromagnets and antiferromagnets.
Recently, a large anomalous Hall effect was reported in the noncollinear antiferromagnetic metal \ce{NbMnP} whose magnetic order is a mixture of the even-parity and the odd-parity magnetic components.
Such a magnetic structure is expected to exhibit the anomalous Hall effect and the nonlinear Hall effect from the symmetry breaking of the antiferromagnet ordering.
Here, we theoretically investigate the intrinsic anomalous and nonlinear Hall effect of \ce{NbMnP} induced by the quantum geometry of Bloch band using the first-principles calculation and the Wannier interpolation method.
We found that the intrinsic Hall response of \ce{NbMnP}  is predominantly governed by the strongly enhanced Berry curvature and Berry-connection-polarization dipole on a specific mirror plane.
These enhanced geometric quantities originate from the spin-orbit-coupling-induced gap openings along the nodal lines.
Our results indicate that \ce{NbMnP} serves as a model system for investigating transport phenomena originating from nodal-lines in parity-mixed antiferromagnets.
\end{abstract}

\maketitle

\section{Introduction}
The anomalous Hall effect (AHE) induces a transverse current in response to an applied electric field in the absence of an external magnetic field. 
Over the past decade, our understanding of the AHE has significantly advanced in the context of Berry phase theory~\cite{RevModPhys.82.1539}. 
This concept has been further extended to related phenomena such as the spin Hall effect~\cite{Murakami_ISH, RevModPhys.87.1213} and the Nernst effect~\cite{Mizuguchi31122019}. 
A key point here is the quantum geometric properties of Bloch wavefunctions, characterized by the quantum geometric tensor~\cite{Provost}.
The imaginary part of this tensor corresponds to the Berry curvature, which induces an anomalous velocity perpendicular to the applied electric field~\cite{RevModPhys.82.1959,Morimoto_JPSJ}.

The AHE has been widely observed in ferromagnets through the interplay between the magnetic moment and spin-orbit coupling (SOC)~\cite{PhysRevLett.88.207208,doi:10.1143/JPSJ.71.19}. 
In recent years, it has been increasingly recognized that the AHE can also emerge in antiferromagnets (AF magnets) which possess no net magnetization~\cite{Smejkal_AHAF}.
AF magnets can exhibit a variety of magnetic structures, including collinear and noncollinear ones. 
Among them, there are the antiferromagnetic states that belong to the same magnetic point group as ferromagnets. 
Interestingly, some of these AF materials exhibit large AHE comparable to that observed in ferromagnets.
Especially, the observation of the large AHE in \ce{Mn3\itshape{Z}} (\ce{\itshape{Z}}=\ce{Sn}, \ce{Ge}) marked a breakthrough in the study of Berry-curvature induced transport properties in AF magnets~\cite{PhysRevLett.112.017205,Kubler_2014,Nakatsuji_Mn3Sn, PhysRevApplied.5.064009,doi:10.1126/sciadv.1501870,PhysRevB.95.094406}. 
Furthermore, the large AHEs have also been reported in materials such as antiferromagnets \ce{CoNb3S6}~\cite{Ghimire_AHE}, \ce{CoTa3S6}~\cite{Takagi_AHE}, and \ce{MnTe}~\cite{PhysRevLett.130.036702}.

\begin{figure}[h]
\centering
\includegraphics[width=1\hsize]{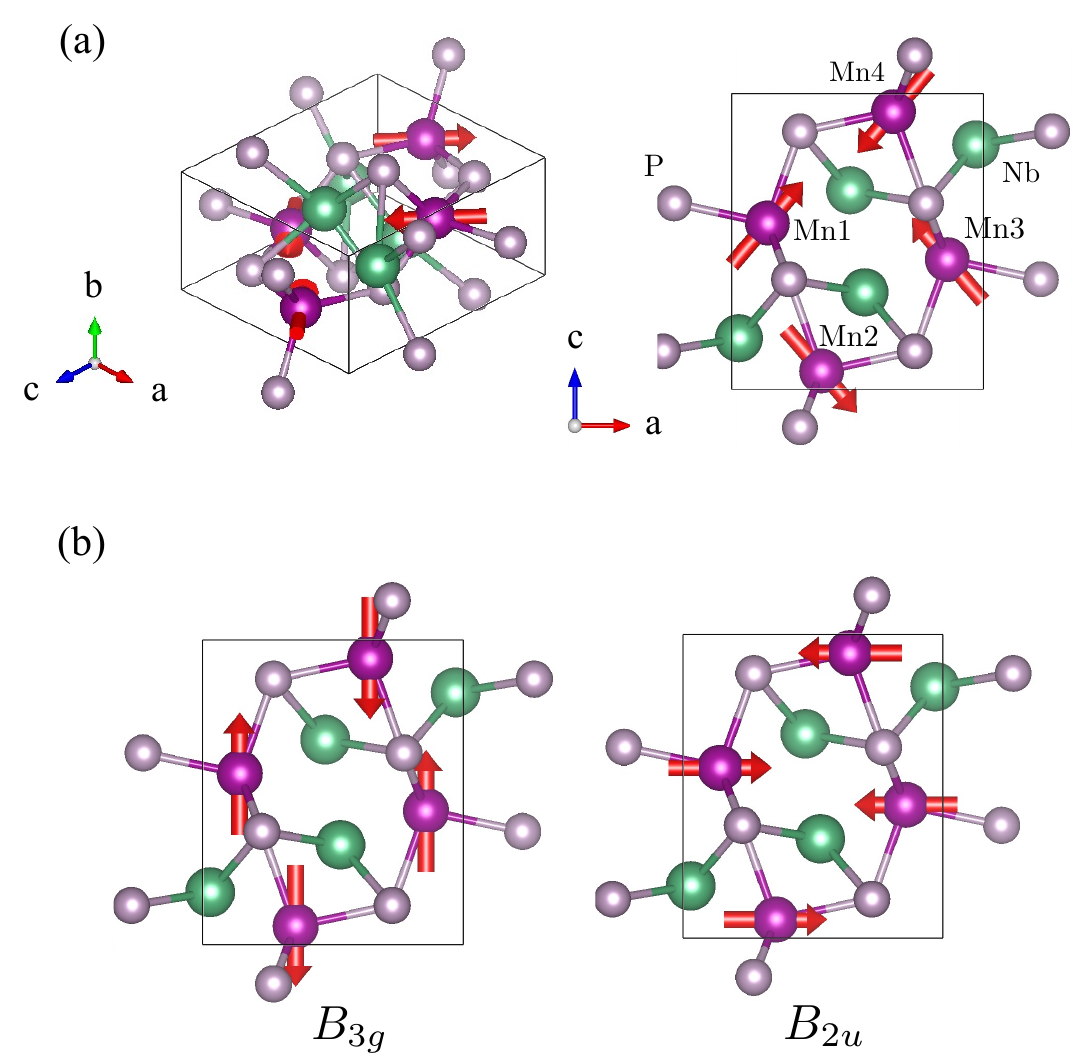}
\caption{(a) Schematic of the \ce{NbMnP} with the noncollinear AF structure. 
(b) Decomposition of the noncollinear AF state into $B_{3g}$ and $B_{2u}$ irreducible representations. The magnetic order in \ce{NbMnP} is a mixture of these two even- and odd-magnetic components. This figure is created by using the \textsc{vesta} software~\cite{Momma:db5098}.}\label{fig:NbMnP_AF}
\end{figure}

Recently, a large AHE has been reported in orthorhombic \ce{NbMnP}~\cite{Kotegawa_NbMnP,doi:10.7566/JPSJ.93.063702}. 
\ce{NbMnP} undergoes a transition to a noncollinear AF metal below its N\'eel temperature $T_N = 233~\rm{K}$.
The magnetic structure of \ce{NbMnP} is experimentally determined by neutron powder diffraction measurements~\cite{PhysRevB.104.174413}, as illustrated in Fig.~\ref{fig:NbMnP_AF}(a).
Four \ce{Mn} atoms in the unit cell have the magnetic moment  $\sim 1.2~\mu_B$ and they form two antiparallel pairs: \ce{Mn}1-\ce{Mn}4 pair and \ce{Mn}2 -\ce{Mn}3 pair, with all moments lying within the $a$-$c$ plane.
The magnetic structure of \ce{NbMnP} consists of the even-parity $B_{3g}$ and odd-parity $B_{2u}$ magnetic component under the crystallographic point group $D_{2h}$ as shown in Fig.~\ref{fig:NbMnP_AF}(b). 
The $B_{3g}$ component belongs to the magnetic space group $Pnm'a'$ (magnetic point group $mm'm'$), which shares the same symmetry as a ferromagnetic structure magnetized along the $a$-axis. 
Consequently, the anomalous Hall conductivity $\sigma_{yz}$ is permitted under the symmetry breaking of magnetic structure with the $B_{3g}$ component~\cite{Kotegawa_NbMnP}, where the $x$-,~$y$- and $z$-axes correspond to the $a$-,~$b$- and $c$-axes, respectively. 

The $B_{2u}$ component belongs to the magnetic space group $Pnm'a$ (magnetic point group $mm'm$) and has an odd parity under both operations of spatial inversion $\mathcal{P}$ and time-reversal $\mathcal{T}$.
Therefore, it preserves $\mathcal{PT}$-symmetry, which leads to forbid the AHE under the pure $B_{2u}$-magnetic structure due to the absence of the Berry curvature.
However, nonlinear Hall effects (NHE) should be generally allowed because the $B_{2u}$-magnetic structure breaks the inversion symmetry. 
In particular, the second-order NHE in AF magnets has recently attracted significant attention~\cite{PhysRevLett.127.277201, PhysRevLett.127.277202, doi:10.1126/science.adf1506, Wang_Nature, PhysRevB.108.L201405}.  
What is important in these studies is that the quantum metric, which corresponds to the real part of the quantum geometric tensor,  plays an essential role and induces the intrinsic NHE~\cite{PhysRevLett.112.166601}.
This Hall effect is in contrast to the dissipative NHE originating from the Berry curvature dipole~\cite{PhysRevLett.115.216806,Ma_Nature7739,KangNatMater4}.

The even-parity $B_{3g}$ and odd-parity $B_{2u}$ magnetic component coexist in \ce{NbMnP} with the noncollinear AF magnetic state, and its magnetic space group is $Pnm'2'_1$ (magnetic point group $mm'2'$).
In this work, we refer to an antiferromagnet in which magnetic components with different parities coexist as ``parity-mixed antiferromagnets"
Due to the symmetry breaking by two magnetic components, the AF states of \ce{NbMnP} is expected to emerge both the AHE and the NHE.
The AHE has already been observed~\cite{Kotegawa_NbMnP}, which supports the presence of the $B_{3g}$ component in \ce{NbMnP}.
In contrast, the evidence that the AF state of \ce{NbMnP} includes the $B_{2u}$ component remains limited, having only been confirmed so far by neutron powder diffraction measurements~\cite{PhysRevB.104.174413}.
The NHE in \ce{NbMnP} is a phenomenon characteristic of the $B_{2u}$-magnetic structure, and this observation provides important evidence supporting the magnetic configuration shown in Fig.~\ref{fig:NbMnP_AF}(a).

In this paper, we present our analysis of the intrinsic anomalous and the nonlinear Hall effects in \ce{NbMnP} to reveal the microscopic origin of the Hall current induced by the quantum geometry of Bloch bands of \ce{NbMnP}.
We firstly perform the first-principles calculation of \ce{NbMnP} with the parity-mixed AF state shown in Fig.~\ref{fig:NbMnP_AF}(a), and constructed the tight-binding model using Wannier interpolation method.
We analyze the Berry curvature and the quantum metric based on the constructed tight-binding model, and discuss the transverse transport properties in \ce{NbMnP} from view point of the symmetry and topology.

\section{Methods}
To investigate the transverse transport properties of \ce{NbMnP}, we consider the electric current expanded by an DC electric field,
\begin{equation}
J_{\alpha}=\sigma_{\alpha\beta}E^\beta+\sigma_{\alpha;\,\beta\gamma}E^\beta E^\gamma,
\end{equation}
where  $\sigma_{\alpha\beta}$ and $\sigma_{\alpha;\,\beta\gamma}$ denote the first- and second-order Hall conductivities, respectively.
Hereafter, we focus on the intrinsic Hall response arising from the quantum geometry of Bloch bands.  
The geometric properties of the Bloch bands are characterized by the quantum geometric tensor defined as
\begin{equation}
Q^{\alpha\beta}_{\bm kn}=\sum_{m(\neq n)}\braket{\partial_\alpha u_{\bm kn}|u_{\bm km}}\braket{u_{\bm km}|\partial_\beta u_{\bm kn}},
\end{equation} 
where $\partial_{\alpha}=\partial/\partial k_{\alpha}$ and the Bloch Hamiltonian $\hat{H}({\bm k})\ket{u_{\bm kn}}=\varepsilon_{\bm kn}\ket{u_{\bm kn}}$ with $\hat{H}(\bm{k})=e^{-i\bm{k}\cdot\hat{\bm{r}}}\hat{H}e^{i\bm{k}\cdot\hat{\bm{r}}}$.
Also, $\ket{u_{\bm kn}}$ represents the periodic part of the Bloch function.
The real and imaginary parts of the quantum geometric tensor, $Q^{\alpha\beta}_{\bm kn}=G^{\alpha\beta}_{\bm kn}-i\Omega^{\alpha\beta}_{\bm kn}/2$, are the quantum metric $G^{\alpha\beta}_{\bm kn} $ and the Berry curvature $\Omega^{\alpha\beta}_{\bm kn}$, respectively. 
Note that the diagonal elements of Berry curvature $\Omega^{\alpha\alpha}_{\bm kn}$ vanish since $Q^{\alpha\alpha}_{\bm kn}$ is a real value. 

The first and second-order intrinsic Hall conductivities have been formulated within both the semiclassical and quantum theoretical framework~\cite{PhysRevLett.112.166601, PhysRevLett.115.216806, PhysRevB.100.165422, Du_NatCommun1}. 
The first-order intrinsic Hall effect corresponds to the Berry-curvature-induced Hall effect i.e. anomalous Hall effect, and this Hall conductivity is written as 
\begin{equation}\label{AHE}
\sigma^{\rm{AHE}}_{\alpha\beta}=-\frac{e^2}{\hbar}\sum_{n}\int\frac{d^3\bm k}{(2\pi)^3}\Omega^{\alpha\beta}_{\bm kn}f(\varepsilon_{\bm kn}-\mu),
\end{equation}
where $f(\varepsilon_{\bm kn}-\mu)$ denotes the Fermi-Dirac distribution function and $\mu$ represents the chemical potential.
The second-order intrinsic Hall effect, called the nonlinear Hall effect (NHE) here, is induced by the Berry-connection polarizability (BCP),
\begin{equation}
\tilde{G}^{\alpha\beta}_{\bm kn}=2\mathrm{Re}\left[\sum_{m(\neq n)}\frac{\braket{\partial_\alpha u_{\bm kn}|u_{\bm km}}\braket{u_{\bm km}|\partial_\beta u_{\bm kn}}}{\varepsilon_{\bm kn}-\varepsilon_{\bm km}}\right],
\end{equation}
which corresponds to the band-normalized quantum metric. 
The nonlinear Hall conductivity is written as 
\begin{equation}\label{NHE}
\sigma^{\rm{NHE}}_{\alpha;\,\beta\gamma}=-\frac{e^3}{\hbar}\sum_{n}\int\frac{d^3\bm k}{(2\pi)^3}\bigg(\partial_{\alpha}\tilde{G}^{\beta\gamma}_{\bm kn}-\partial_{\beta}\tilde{G}^{\alpha\gamma}_{\bm kn}\bigg)~f(\varepsilon_{\bm kn}-\mu).
\end{equation}

To calculate the intrinsic Hall conductivities~\eqref{AHE} and \eqref{NHE}, we constructed a tight-binding model from the first-principles band structure
(see Appendix~\ref{App.A} section for details). 
The electronic structure of \ce{NbMnP} was obtained from the first-principles calculations performed using the Quantum Espresso package~\cite{Giannozzi_2009}.
Here, spin-orbit coupling is  included in our calculation.
Exchange correlation energy was approximated using the generalized gradient approximation (GGA) as parametrized by  Perdew, Burke, and Ernzerhof~\cite{PhysRevLett.77.3865}. 
The pseudopotentials in the projector augmented-wave method were provided by the PSLibrary~\cite{PhysRevB.50.17953,PhysRevB.59.1758,DALCORSO2014337}.
The lattice constants were taken from the experimental values at $T=\qty{9}{\kelvin}$, $a=\qty{6.1661}{\angstrom}$, $b=\qty{3.5325}{\angstrom}$, $c=\qty{7.2199}{\angstrom}$~\cite{PhysRevB.104.174413}.
The atomic positions were relaxed starting from the experimental atomic positions, until the residual forces were less than $0.01~[\mathrm{eV/\AA}]$. 
The cut-off energies for the plane-wave basis set and charge density were set to $\qty{50}{\rydberg}$ and $\qty{400}{\rydberg}$, respectively. 
A $k$-mesh of $9 \times 15 \times 9$ was used in the first Brillouin zone, with a Methfessel-Paxton smearing width of $\qty{0.005}{\rydberg}$ to determine the Fermi level.

\begin{figure}[h]
\centering
\includegraphics[width=1\hsize]{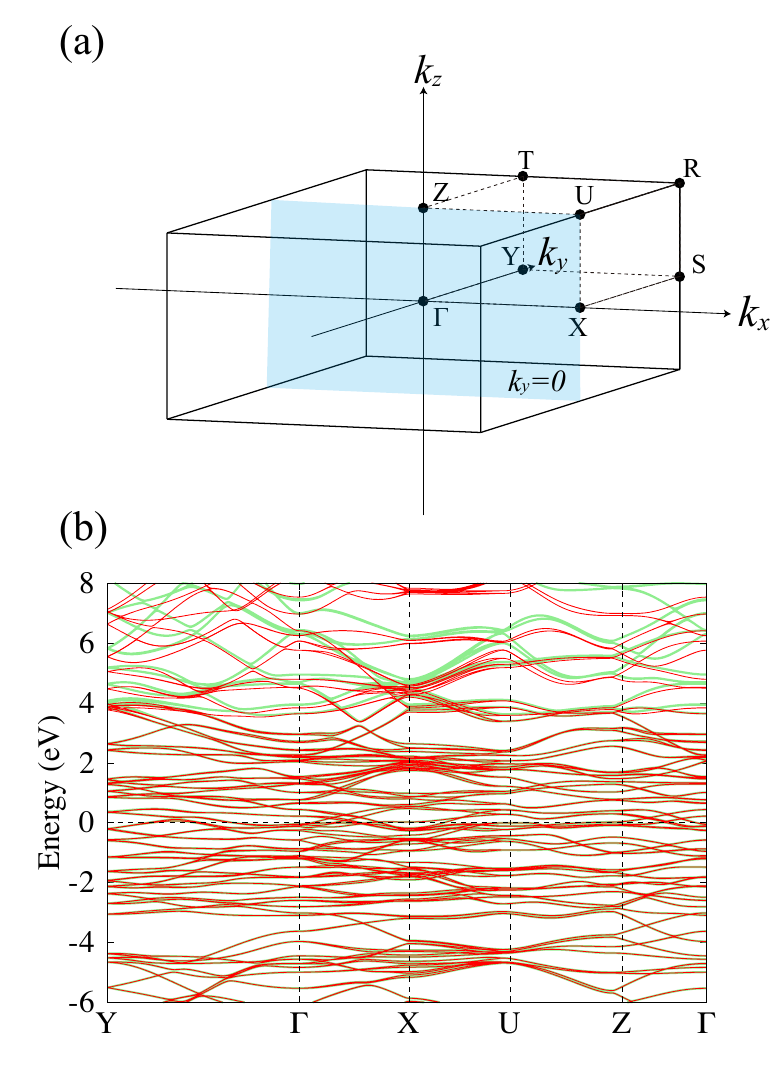}
\caption{(a) Magnetic-Brillouin zone for the orthorhombic lattice of \ce{NbMnP}. Area painted in light blue represents the $k_y=0$ plane. (b) Energy bands of \ce{NbMnP} in the noncollinear AF magnetic state from the first-principles calculation (red lines) and the tight-binding model (green lines).}\label{fig:Band}
\end{figure}

The tight-binding model was constructed from the $4d$ and $5s$ orbitals of \ce{Nb}, $3d$ and $4s$ orbitals of \ce{Mn}, $3s$ and $3p$ orbitals of \ce{P} as projection functions, using Wannier90 package~\cite{Pizzi_2020}.
Figure~\ref{fig:Band}(b) shows the band structure of \ce{NbMnP} under the noncollinear AF magnetic structure.
The red lines in Fig.~\ref{fig:Band}(b) show the electronic band structure obtained from first-principles calculations, whereas the green lines show the bands obtained from the tight-binding model, which accurately reproduce the first-principles band dispersion from well below the Fermi level up to about $\qty{4}{\electronvolt}$ above it.

\section{Anomalous Hall effect}
We here analyze the anomalous Hall conductivity $\sigma_{\alpha\beta}$ in \ce{NbMnP} at \qty{100}{\kelvin}. 
The noncollinear AF magnetic state of $\ce{NbMnP}$ is expressed by a linear combination of the even-parity $B_{3g}$ and odd-parity $B_{2u}$ irreducible representations~\cite{Kotegawa_NbMnP}.
Such the parity-mixed AF magnetic state triggers the emergence of both the anomalous and nonlinear Hall effect.
We summarize character of the irreducible representations $B_{3g}$ and $B_{2u}$ for each symmetry operation of the point group of \ce{NbMnP} in Table~\ref{tab:symmetry}.
This table serves as the basis for determining the allowed components of the linear and nonlinear Hall conductivities. 
Each symmetry operator has the character $+1$ and $-1$ for the $B_{3g}$ and $B_{2u}$ magnetic structures, where $+1$ indicates that the magnetic configuration is invariant under the operation, while $-1$ means that the configuration becomes invariant only after applying time reversal. 
Since the magnetic structure in Fig.~\ref{fig:NbMnP_AF}(a) can be regarded as a linear combination of the $B_{3g}$ and $B_{2u}$ magnetic configurations, those symmetry operations that leave both configurations invariant also leave the combined magnetic structure invariant.
From Table~\ref{tab:symmetry}, the magnetic structure illustrated in Fig.~\ref{fig:NbMnP_AF}(a) is invariant under the symmetry operators $E,~\sigma_x,~\mathcal TC_{2z},~\mathcal T\sigma_{y}$, where $\mathcal T$ represents the time-reversal operator.

\begin{table*}[htbp]
\caption{\label{tab:symmetry}
Character table of the $B_{3g}$ and $B_{2u}$ irreducible representations for the crystallographic point group $D_{2h}$ of \ce{NbMnP}. 
Here, $E$ denotes the identity, $C_{2x}$, $C_{2y}$, and $C_{2z}$ are twofold rotations about the Cartesian axes, 
$I$ is inversion, and $\sigma_{i=x,y,z}$ is mirror reflections with respect to the 
$i=0$ plane.  A character of $-1$ indicates that the operation leaves the 
system invariant only when combined with time-reversal symmetry. }
\begin{ruledtabular}\label{MSG}
\begin{tabular}{c|cccccccc}
      & $E$ & $C_{2x}$ & $C_{2y}$ & $C_{2z}$ & $I$ & $\sigma_x$ & $\sigma_y$ & $\sigma_z$ \\
\hline
$B_{3g}$ & $1$ &  $1$   &  $-1$    &    $-1$     &  $1$  &  $1$ &  $-1$ &   $-1$  \\
$B_{2u}$ & $1$ &  $-1$   &    $1$     &  $-1$    &  $-1$  &  $1$ &   $-1$   &  $1$ \\
\end{tabular}
\end{ruledtabular}
\end{table*}

\begin{figure}[h]
\centering
\includegraphics[width=1\hsize]{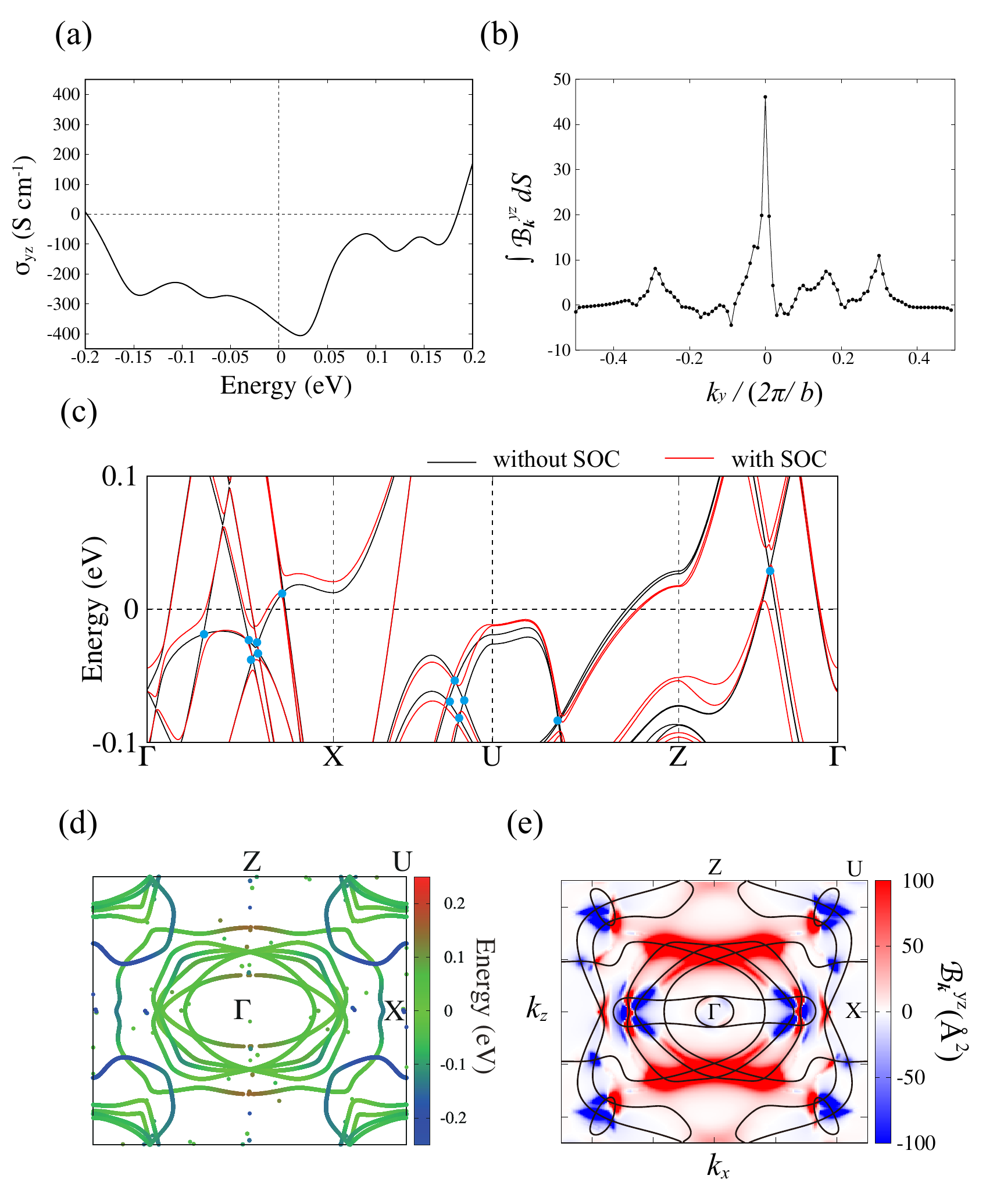}
\caption{
(a) Chemical potential dependence of the anomalous Hall conductivity $\sigma_{yz}$. 
Fermi level is set to zero on the horizontal axis. 
(b) $k_y$-dependence of area integral of $\mathcal{B}^{yz}_{\bm k}$ over the $k_x$-$k_z$ plane. 
(c) Electronic bands of \ce{NbMnP} with and without SOC. 
(d) Nodal lines on the $k_y=0$ plane. The line color represents the energy level of the nodal lines. 
(e) Berry curvature summed over the occupied states, $\mathcal B^{yz}_{\bm k}$, on the $k_y=0$ plane. 
Black lines represent the Fermi surface. }\label{fig:topo}
\end{figure}

The symmetry of \ce{NbMnP} allows the anomalous Hall conductivity $\sigma^{\rm{AHE}}_{yz}$, which have been experimentally observed~\cite{Kotegawa_NbMnP}.
Note that the Berry curvature satisfies $\Omega^{\alpha\beta}_{\bm kn}=-\Omega^{\alpha\beta}_{\bm kn}$ i.e. $\Omega^{\alpha\beta}_{\bm kn}= 0$ under the $B_{2u}$-magnetic symmetry.
Therefore, the symmetry breaking by $B_{3g}$-AF magnetic component directly triggers this AHE.
Figure \ref{fig:topo}(a) shows the chemical potential dependence of the anomalous Hall conductivity $\sigma^{\rm{AHE}}_{yz}$ calculated based on the tight-binding model. 
The anomalous Hall conductivity is enhanced near the Fermi level, reaching $\qty[per-mode = symbol]{-366}{\siemens\per\centi\metre}$, which roughly agrees with the experimental value.
This large AHE remains robust against shifts of the chemical potential in the range of $-0.15$ to $\qty{0.15}{\electronvolt}$.

To investigate the origin of the large AHE in \ce{NbMnP}, we introduce the Berry curvature summed over the occupied states,
\begin{equation}\label{BC_sum}
\mathcal{B}^{\alpha\beta}_{\bm k}=\sum_{n}\Omega^{\alpha\beta}_{\bm kn}f(\varepsilon_{\bm kn}-\mu).
\end{equation}
The momentum-average of $\mathcal{B}^{\alpha\beta}_{\bm k}$ corresponds to the anomalous Hall conductivity $\sigma^{\rm{AHE}}_{\alpha\beta}$ from Eq.~\eqref{AHE}.
Figure.~\ref{fig:topo}(b) shows the $k_y$-dependence of the area integral of $\mathcal{B}^{yz}_{\bm k}$ over the $k_x$-$k_z$ plane.
Here, chemical potential is set to the Fermi level.
Interestingly, the value of $\int\mathcal{B}^{yz}_{\bm k}dS$ at $k_y=0$ is much larger than that at $k_y\neq 0$. 
To evaluate the contribution of the pronounced peak around $k_y=0$ to the $\sigma^{\rm{AHE}}_{yz}/(-e^2/h)$, we perform the $k_y$-integration of the function shown in Fig.~\ref{fig:topo}(b), restricting the integration to contributions satisfying $\int\mathcal{B}^{yz}_{\bm k}dS \geq 15$. 
We find that this integral accounts for approximately $35$\% of the anomalous Hall conductivity $\sigma^{\rm{AHE}}_{yz}$.

The enhancement of the area integral at the $k_y=0$ plane is related to the topological nature of Bloch bands.
In the absence of the SOC, the magnetic structure of \ce{NbMnP} is invariant under the $\mathcal{U}\sigma_y$, where $\mathcal{U}$ denotes the operator that rotates the spin by \ang{180} along the $y$-axis.
This implies that the magnetic state of NbMnP without SOC effectively possesses a $y$-mirror symmetry, 
and symmetry-protected band degeneracies appear on the corresponding mirror plane. 
The $k_y=0$ plane, where the Berry curvature is strongly enhanced, can therefore be regarded as one of the $y$-mirror planes.
To see this, we analyze the topological degeneracies in the electronic bands of \ce{NbMnP}.
Figure~\ref{fig:topo}(c) shows the electronic bands of \ce{NbMnP} with and without SOC.
We select the $k$-path connecting the high-symmetry points in the $k_y=0$ plane shown in Fig.~\ref{fig:Band}(a).
The blue points represent the topological degeneracies in the electronic structure without SOC.
These degeneracies are expected to be topologically protected by the effective $y$-mirror symmetry and they form the nodal line on the $y$-mirror plane as shown in Fig.~\ref{fig:topo}(d). 
While the nodal lines usually appear on the mirror plane as the intersections of the energy bands with opposite eigenvalues of the ordinary mirror symmetry~\cite{Vanderbilt_book_2018}, the effective $y$-mirror symmetry  $\mathcal{U}\sigma_y$ produces the nodal lines on the $k_y=0$ plane in the present case despite the absence of the ordinary $y$-mirror symmetry $\sigma_y$.
The line color in Fig.~\ref{fig:topo}(d) represents the energy level of the nodal lines, which shows that the nodal lines are located near the Fermi energy.

The nodal lines are lifted by SOC as shown in Fig.~\ref{fig:topo}(c), leading to the formation of a small band gap.
This gap opening strongly enhances the Berry curvature along the nodal lines in the $k_y=0$ plane.
Figure.~\ref{fig:topo}(e) shows the momentum-resolved Berry curvature $\mathcal B^{yz}_{\bm k}$ in the $k_y=0$ plane.
The corresponding plane is illustrated in Fig.~\ref{fig:Band}(a).
The black lines represent the Fermi surface.
The Berry curvature takes predominately positive values over the $k_y=0$ plane, and is strongly enhanced in the regions where the nodal lines are densely distributed, in particular. 
Although the magnetic structure of \ce{NbMnP} is expressed as the linear combination of the $B_{3g}$- and $B_{2u}$-AF magnetic components, the origin of the Berry curvature $\mathcal B^{yz}_{\bm k}$ is directly associated with the symmetry breaking of $B_{3g}$ magnetic component.
The Berry curvature $\mathcal B^{yz}_{\bm k}$ is symmetric with respect to the mirror reflection $\sigma_y$ combined with time-reversal symmetry i.e. $\mathcal B^{yz}(-k_x,k_y,-k_z)=\mathcal B^{yz}(k_x,k_y,k_z)$.  
To investigate the origin of the peak as shown in Fig.~\ref{fig:topo}(b),
we perform the are integral of $\mathcal{B}^{yz}_{\bm k}$ over the $k_y=0$ plane, restricting the integration to contributions satisfying $|\mathcal B^{yz}_{\bm k}| > \qty{50}{\square\angstrom}$. 
As a result, we find that this integral accounts for about $94$\% of the pronounced peak at $k_y=0$ shown in Fig.~\ref{fig:topo}(b).
These results demonstrate that the Berry curvature enhanced by the gap opening along the nodal lines leads to the emergence of the large AHE.

In the noncollinear AF magnet \ce{Mn3{\itshape{A}}N} known to exhibit the large AHE, the cumulative integration of the small Berry curvature distributed over the Brillouin zone mainly contributes to the anomalous Hall conductivity~\cite{PhysRevB.100.094426}.  
In \ce{NbMnP}, as in \ce{Mn3{\itshape{A}}N}, the small Berry-curvature contribution to the Hall conductivity is appreciable; at the same time, a major contribution arises from the strongly enhanced Berry curvature in the vicinity of the nodal lines, which is characteristic of \ce{NbMnP}.
Our results shows that the strongly enhanced Berry curvature exceeding $\qty{50}{\square\angstrom}$ on the particular mirror plane also leads to the large AHE in \ce{NbMnP}.

\section{Nonlinear Hall effect}
Here, we analyze the nonlinear Hall conductivity~\eqref{NHE} in \ce{NbMnP}. 
Because the current response proportional to $E^\beta E^\gamma$ is forbidden in centrosymmetric systems, this Hall effect arises from the symmetry breaking induced by the $B_{2u}$-AF magnetic component.
To extract a finite component of nonlinear conductivity tensor, we introduce the rank-$2$ pseudotensor
\begin{equation}
\sigma_{\gamma\delta}=\mathcal{E}_{\alpha\beta\gamma}\frac{\sigma^{\rm{NHE}}_{\alpha;\,\beta\delta}}{2}
\end{equation}
reflecting the antisymmetry $\sigma^{\rm{NHE}}_{\alpha;\,\beta\gamma}=-\sigma^{\rm{NHE}}_{\beta;\,\alpha\gamma}$.
Here, $\mathcal{E}_{\alpha\beta\gamma}$ is the Levi-Civita symbol.
The rank-$2$ pseudotensor $\sigma$ is constrained by
\begin{equation}
\sigma=\eta_T\det(\,\mathcal R\,)\mathcal R~\sigma~\mathcal R^{-1},
\end{equation}
where $\mathcal R$ is a point group operation.
The factor $\eta_T$ becomes $-1$ for anti-unitary operations ($\mathcal T\mathcal O$) and $ +1$ for unitary operations ($\mathcal O$) when $\sigma$ is $\mathcal T$-odd.
Here, $\mathcal O$ is a spatial operation.
Based on the magnetic point group symmetry of \ce{NbMnP} 
as shown in Table~\ref{tab:symmetry}, the finite nonlinear Hall conductivities in \ce{NbMnP} are $\sigma^{\rm{NHE}}_{y;\,xx},~\sigma^{\rm{NHE}}_{y;\,zz},~\sigma^{\rm{NHE}}_{x;\,yx}$ and $\sigma^{\rm{NHE}}_{z;\,yz}$. 
Among them, we focus on the components of the nonlinear conductivity tensor $\sigma^{\rm{NHE}}_{y;\,xx}$ and $\sigma^{\rm{NHE}}_{y;\,zz}$ to discuss the DC current responses.

\begin{figure}[t]
\centering
\includegraphics[width=1\hsize]{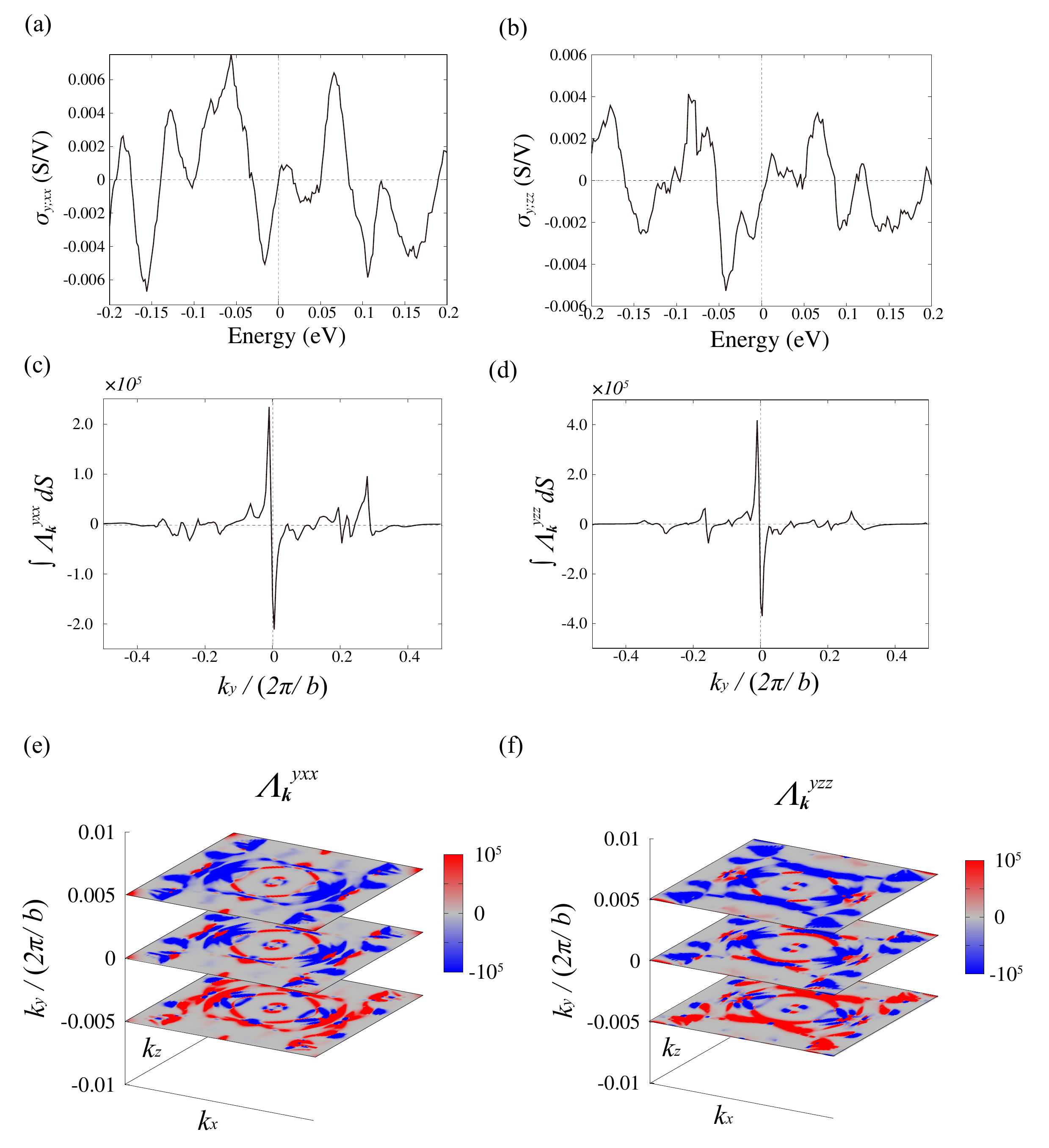}
\caption{(a)(b) Chemical potential dependence of the intrinsic nonlinear Hall conductivity $\sigma^{\rm{NHE}}_{y;\,xx}$ and  $\sigma^{\rm{NHE}}_{y;\,zz}$, respectively. 
(c)(d) Area integral of $\Lambda^{\alpha\beta\gamma}_{\bm k}$ over the $k_x$-$k_z$ plane. 
(e)(f) BCP dipole summed over the occupied states, $\Lambda^{yxx}_{\bm k}$ and $\Lambda^{yzz}_{\bm k}$, on the $k_y=-0.005(2\pi/b),~0,~0.005(2\pi/b)$ planes. Black lines represent the Fermi surface.}\label{fig:INHE}
\end{figure}

We show the chemical potential dependence of the nonlinear Hall conductivity $\sigma^{\rm{NHE}}_{\alpha;\,\beta\beta}$ in Fig.~\ref{fig:INHE} (a) and (b). 
The temperature is set to $\qty{100}{\kelvin}$. 
In contrast to the AHE, the nonlinear conductivity drastically changes with variations in the chemical potential.
DC Hall conductivities $\sigma^{\rm{NHE}}_{y;\,xx}$ and $\sigma^{\rm{NHE}}_{y;\,zz}$ reach $-0.12$ and $\qty[per-mode = symbol]{-0.95}{\milli\siemens\per\volt}$, respectively, when the chemical potential equals the Fermi level. 
The nonlinear Hall conductivity is given by the momentum average of the Berry connection polarization (BCP) dipole summed over the occupied states,
\begin{equation}\label{BCP_sum}
\Lambda^{\alpha\beta\gamma}_{\bm k}=-\sum_n\bigg(\partial_{\alpha}\tilde{G}^{\beta\gamma}_{\bm kn}-\partial_{\beta}\tilde{G}^{\alpha\gamma}_{\bm kn}\bigg)~f(\varepsilon_{\bm kn}-\mu).
\end{equation}
Figure~\ref{fig:INHE} (c) and (d) show the $k_y$-dependence of the area integrals of $\Lambda^{yxx}_{\bm k}$ and $\Lambda^{yzz}_{\bm k}$ over the $k_x$-$k_z$ plane.
Both area integrals exhibit the sharp peak component with opposite signs across the plane $k_y\sim -0.003(2\pi/b)$.
The positions of two peaks are located at $k_y=-0.01(2\pi/b),~0.005(2\pi/b)$.
We evaluate the $k_y$-integrals of the function shown in Fig.~\ref{fig:INHE}(c) and (d), restricting the integration to the contributions satisfying 
$|\int \Lambda_{\bm k}^{y\alpha\alpha}dS| > \qty{e5}{\per\electronvolt\per\angstrom}$. 
As a result, we find that these integrals are essentially dominated only by the sharp peak component in the vicinity of the $k_y=0$ plane and account for about $45$\% of each nonlinear Hall conductivity.
In the AHE,  the pronounced peak near the $k_y=0$ plane directly contributes to the anomalous Hall conductivity $\sigma^{\rm{AHE}}_{yz}$, as shown in Fig.~\ref{fig:topo} (b).
In the NHE, on the other hand,  the incomplete cancellation of the two pronounced peaks with opposite sign around $k_y=0$ plane makes a dominant contribution to the NHE, as shown in Fig.~\ref{fig:INHE} (c) and (d).

We plot the momentum-resolved BCP dipole for each $k_x$-$k_z$ plane in Fig.~\ref{fig:INHE} (e) and (f). 
The BCP dipole is symmetric under the action of $\mathcal T\sigma_y$ operator, that is $\Lambda^{y\alpha\alpha}(k_x,k_y,k_z)=\Lambda^{y\alpha\alpha}(-k_x,k_y,-k_z)~(\alpha=x,~z)$.
In near the $k_y=0$ plane, the BCP dipole is strikingly enhanced in the regions where the nodal lines are densely distributed, similar to the Berry curvature (see Fig.~\ref{fig:topo}(d) and (e)).
The BCP dipole takes predominately negative values over the $k_y=0.005(2\pi/b)$ plane while takes predominately positive values over the $k_y=-0.005(2\pi/b)$ plane.
This $k_y$-dependence of  the momentum-resolved BCP dipole makes the two pronounced peaks with opposite sign in the Fig.~\ref{fig:INHE}(c) and (d), which suggests that the BCP dipole enhanced by the SOC-induced gap opening along the nodal lines dominates the emergence of the NHE.

Considering both the AHE and the NHE, we find that the Hall current response depends on the direction of the applied electric field.
When an electric field is applied along the x axis, the Hall current, $J_y=\sigma^{\mathrm{NHE}}_{y;\,xx}E_x^2$, flows along the y axis.
This Hall current retains the same direction even when the applied electric field is reversed.
In contrast, when an electric field is applied along the z axis, the Hall current, $J_y=(\sigma^{\mathrm{AHE}}_{yz}+\sigma^{\mathrm{NHE}}_{y;\,zz}E_z)E_z$, is induced.
This Hall response is asymmetric under the reversal of the applied electric field.
Such a difference in the directional Hall response is not observed for the pure $B_{2u}$-magnetic component, but emerges when the $B_{2u}$-magnetic component coexists with the $B_{3g}$-magnetic component. 

\section{Conclusion}
In this work, we constructed the tight-binding model of the noncollinear AF metal \ce{NbMnP} from the first-principles electronic bands and analyzed the AHE and the NHE from view point of the symmetry and the quantum geometric nature of Bloch function.
As a result, we revealed that the gap opening of the nodal lines on the $k_y=0$ plane enhances the quantum geometric quantities, namely the Berry curvature and the BCP dipole, thereby leading to the emergence of both the AHE and the NHE in \ce{NbMnP}.
Firstly, the AHE is induced by the symmetry breaking arising from the even-parity $B_{3g}$ magnetic component, and the anomalous Hall conductivity $\sigma_{yz}$ reaches $-366~\rm{S/cm}$ at the Fermi level.
This AHE is enhanced by the Berry curvature summed over the occupied states on the $k_y=0$ plane.
This Berry curvature on the $y$-mirror plane takes a large value in the region where the nodal lines gapped by the SOC are densely distributed, in the momentum space.
Nextly, the NHE is induced by the symmetry breaking arising from the odd-parity $B_{2u}$ magnetic component.
By the symmetry analysis,  we showed that the nonlinear Hall responses $\sigma_{y;xx}$ and $\sigma_{y;zz}$ are allowed as DC response.
Both nonlinear Hall conductivities reach the order of a few $~[\mathrm{mS/V}]$ at the Fermi level.
The dominant contribution to the NHE arises from the strongly enhanced BCP dipole summed over the occupied states near the $y$-mirror plane.
We have focused on the intrinsic contribution to the Hall current to investigate the Hall response arising from the parity-mixed magnetic state.
By contrast, the dissipative Hall response, which sensitively depends on scattering mechanisms and relaxation processes, is left for future investigation in NbMnP.
In summary, \ce{NbMnP} exhibits both the Berry-curvature-driven AHE and the Berry-connection-polarization-driven NHE owing to the coexistence of odd-parity and even-parity magnetic components. 
Our results show that \ce{NbMnP} provides a prototypical platform for exploring transport properties emerging from the nodal lines of the parity-mixed antiferromagnets.

\begin{acknowledgments}
We thank H. Kotegawa and H. Ikeda for helpful comments and discussions.
We are also grateful to K. Tanno for the technical supports. 
This research is supported by JSPS KAKENHI Grants Numbers JP23H01130, JP23K20824, JP24K00581, JP24K00588, JP25K00947, JP25K21684.
We also acknowledge the use of supercomputing system, MASAMUNE-IMR, at CCMS, IMR, Tohoku University in Japan.
\end{acknowledgments}

\appendix
\section{The Berry curvature (BC) and Berry connection polarization (BCP) dipole by Wannier interpolation method}\label{App.A}
Here, we evaluate the Berry curvature~\eqref{BC_sum} and Berry connection polarization dipole~\eqref{BCP_sum} summed over the occupied states by Wannier interpolation method.
We introduce the Bloch-like functions given by
\begin{equation}
\ket{\tilde u_{\bm kn}}=\frac{1}{\sqrt{N}}\sum_{\bm R}e^{-i\bm k\cdot(\hat{\bm{r}}-\bm R)}\ket{w_{\bm Rn}},
\end{equation}
where $\ket{w_{\bm Rn}}$ is the Wannier functions localized at unit cell $\bm R$ and $N$ denotes the total number of unit cells.
The Wannier functions are constructed from a set of original Bloch functions $\ket{u_{\bm kn}}$ of interest.
The number of Bloch-like functions $\ket{\tilde u_{\bm kn}}$ must be larger than that one of the original Bloch functions $\ket{u_{\bm kn}}$.
The tight-binding Hamiltonian based on the Wannier interpolation method is written as 
\begin{align}
    \tilde{\mathcal H}_{\bm k,nm}=\braket{\tilde u_{\bm kn}|\hat H(\bm{k})|\tilde u_{\bm km}} =\sum_{\bm R}e^{-i\bm k\cdot \bm R}\braket{w_{\bm Rn}|\hat H|w_{\bm 0m}}.
\end{align}

We introduce the unitary transformation matrix $U_{\bm k}$ such as
\begin{equation}
\bar{\mathcal{H}}_{\bm k,nm}=U^\dag_{\bm k}\tilde{\mathcal{H}_{\bm k}}U_{\bm k}=\bar{\varepsilon}_{\bm kn}\delta_{nm}.
\end{equation}
Note that $\bar{\varepsilon}_{\bm kn}$ maches to the occupied actual Bloch bands $\varepsilon_{\bm kn}$. The corresponding Bloch functions are obtained by the unitary transformation,
\begin{equation}
\ket{\bar u_{\bm kn}}=\sum_m\ket{\tilde u_{\bm km}}U_{\bm k,mn}.
\end{equation}
Similar to $\bar{\varepsilon}_{\bm kn}$, $\ket{\bar u_{\bm kn}}$ also is identical to the occupied actual Bloch function  $\ket{u_{\bm kn}}$.

To caluculate the BC and BCP dipole summed over the occupied states, we can use the eigenstates $\ket{\bar u_{\bm kn}}$ instead of the occupied actual Bloch function  $\ket{u_{\bm kn}}$; therefore we consider the BC and BCP in the $\{\ket{\bar u_{\bm kn}}\}$ basis,
\begin{align}
&\bar{\Omega}^{\alpha\beta}_{\bm kn}=-2\mathrm{Im}\left[\sum_{m(\neq n)}\bar{A}^{\alpha}_{\bm k,nm}\bar{A}^{\beta}_{\bm k,mn}\right],\\
&\bar{G}^{\alpha\beta}_{\bm kn}=2\mathrm{Re}\left[\sum_{m(\neq n)}\frac{\bar{A}^{\alpha}_{\bm k,nm}\bar{A}^{\beta}_{\bm k,mn}}{\bar\varepsilon_{\bm kn}-\bar\varepsilon_{\bm km}}\right].
\end{align}
Here, $\bar{A}^{\alpha}_{\bm k,nm}=i\braket{\bar u_{\bm kn}|\partial_{\alpha}\bar u_{\bm km}}$
is the Berry connection matrix in the multi-band systems. 
The derivative of $\ket{\bar u_{\bm kn}}$ is written as 
\begin{equation}
\ket{\partial_{\alpha}\bar u_{\bm kn}}=\sum_{m}\ket{\partial_{\alpha}\tilde u_{\bm km}}U_{\bm k,mn}
+\sum_{m}\ket{\bar u_{\bm km}}\bar{\mathcal D}^\alpha_{\bm k,mn}, 
\end{equation}
where $\bar{\mathcal D}^\alpha_{\bm k} = U^\dag_{\bm k}\partial_{\alpha}U_{\bm k}$ is referred to as the $\mathcal D$-matrix. 
Here, we choice the gauge such as the diagonal components of $\mathcal D$-matrix are zero, and the $\mathcal D$-matrix is written as
\begin{equation}
\bar{\mathcal D}^\alpha_{\bm k,nm} =
\begin{cases}
    0& n=m \\
    -\frac{\bar{\mathcal{V}}^{\alpha}_{\bm k,nm}}{\bar{\varepsilon}_{\bm kn}-\bar{\varepsilon}_{\bm km}}& n\neq m,
  \end{cases}
\end{equation} 
where $\bar{\mathcal{V}}^{\alpha}_{\bm k}=U^\dag_{\bm k}\partial_{\alpha}\tilde{\mathcal{H}}_{\bm k}U_{\bm k}$.
Then, the Berry connection matrix is written as
\begin{equation}
\bar{A}^{\alpha}_{\bm k,nm}=\bar{\mathcal A}^{\alpha}_{\bm k,nm} + i\bar{\mathcal{D}}_{\bm k,nm}^\alpha,
\end{equation}
where $\bar{\mathcal A}^{\alpha}_{\bm k}=U^\dag_{\bm k}\tilde{A}_{\bm k}U_{\bm k}$ and 
$\tilde{A}_{\bm k,nm}=i\braket{\tilde u_{\bm kn}|\partial_{\alpha}\tilde u_{\bm km}}$. 

Consequently, the BC and BCP summed over the occupied state are written as
\widetext
\begin{align}
\label{BC_sum_A}
&\sum_n\bar{\Omega}^{\alpha\beta}_{\bm kn}f_{\bm kn}=\sum_{n,m}\bigg(
i\bar{\mathcal A}^{\alpha}_{\bm k,nm}\bar{\mathcal A}^{\beta}_{\bm k,mn}
+\bar{\mathcal{D}}_{\bm k,nm}^\beta\bar{\mathcal A}^{\alpha}_{\bm k,mn}
-\bar{\mathcal{D}}_{\bm k,nm}^\alpha\bar{\mathcal A}^{\beta}_{\bm k,mn}
-i\bar{\mathcal{D}}_{\bm k,nm}^\alpha\bar{\mathcal{D}}_{\bm k,mn}^\beta
\bigg)
\bigg(
f_{\bm kn}-f_{\bm km}
\bigg),\\
\label{BCP_sum_A}
&\sum_n\bar{G}^{\alpha\beta}_{\bm kn}f_{\bm kn}=\sum_{n,m}\bigg(
\frac{\bar{\mathcal A}^{\alpha}_{\bm k,nm}\bar{\mathcal A}^{\beta}_{\bm k,mn}}{\bar\varepsilon_{\bm kn}-\bar\varepsilon_{\bm km}}
+i\frac{\bar{\mathcal{D}}_{\bm k,nm}^\beta\bar{\mathcal A}^{\alpha}_{\bm k,mn}}{\bar\varepsilon_{\bm kn}-\bar\varepsilon_{\bm km}}
+i\frac{\bar{\mathcal{D}}_{\bm k,nm}^\alpha\bar{\mathcal A}^{\beta}_{\bm k,mn}}{\bar\varepsilon_{\bm kn}-\bar\varepsilon_{\bm km}}
-\frac{\bar{\mathcal{D}}_{\bm k,nm}^\alpha\bar{\mathcal{D}}_{\bm k,mn}^\beta}{\bar\varepsilon_{\bm kn}-\bar\varepsilon_{\bm km}}
\bigg)
\bigg(
f_{\bm kn}-f_{\bm km}
\bigg),
\end{align}
\endwidetext
\noindent respectively, where $f_{\bm kn}=f(\bar{\varepsilon}_{\bm kn}-\mu)$. 
The terms arising from pairs of teh occupied states in Eqs~\eqref{BC_sum_A} and \eqref{BCP_sum_A} make no contribution due to the factor $f_{\bm kn}-f_{\bm km}$. 
The fourth term, called $\mathcal D$-$\mathcal D$ term, cantains the highest-order energy denominator among all the contributions and gives the dominant contribution.
Therefore, we drop the first-third term in Eqs~\eqref{BC_sum_A} and \eqref{BCP_sum_A} and calculate
\begin{align}
&\sum_n\bar{\Omega}^{\alpha\beta}_{\bm kn}f_{\bm kn}
=-i\sum_{n,m}\bar{\mathcal{D}}_{\bm k,nm}^\alpha\bar{\mathcal{D}}_{\bm k,mn}^\beta
\bigg(
f_{\bm kn}-f_{\bm km}
\bigg),\\
&\sum_n\bar{G}^{\alpha\beta}_{\bm kn}f_{\bm kn}
=-\sum_{n,m}\frac{\bar{\mathcal{D}}_{\bm k,nm}^\alpha\bar{\mathcal{D}}_{\bm k,mn}^\beta}{\bar\varepsilon_{\bm kn}-\bar\varepsilon_{\bm km}}
\bigg(
f_{\bm kn}-f_{\bm km}
\bigg).
\end{align}

\newpage
\bibliography{apssamp}

\begin{thebibliography}{42}%
\makeatletter
\providecommand \@ifxundefined [1]{%
 \@ifx{#1\undefined}
}%
\providecommand \@ifnum [1]{%
 \ifnum #1\expandafter \@firstoftwo
 \else \expandafter \@secondoftwo
 \fi
}%
\providecommand \@ifx [1]{%
 \ifx #1\expandafter \@firstoftwo
 \else \expandafter \@secondoftwo
 \fi
}%
\providecommand \natexlab [1]{#1}%
\providecommand \enquote  [1]{``#1''}%
\providecommand \bibnamefont  [1]{#1}%
\providecommand \bibfnamefont [1]{#1}%
\providecommand \citenamefont [1]{#1}%
\providecommand \href@noop [0]{\@secondoftwo}%
\providecommand \href [0]{\begingroup \@sanitize@url \@href}%
\providecommand \@href[1]{\@@startlink{#1}\@@href}%
\providecommand \@@href[1]{\endgroup#1\@@endlink}%
\providecommand \@sanitize@url [0]{\catcode `\\12\catcode `\$12\catcode `\&12\catcode `\#12\catcode `\^12\catcode `\_12\catcode `\%12\relax}%
\providecommand \@@startlink[1]{}%
\providecommand \@@endlink[0]{}%
\providecommand \url  [0]{\begingroup\@sanitize@url \@url }%
\providecommand \@url [1]{\endgroup\@href {#1}{\urlprefix }}%
\providecommand \urlprefix  [0]{URL }%
\providecommand \Eprint [0]{\href }%
\providecommand \doibase [0]{https://doi.org/}%
\providecommand \selectlanguage [0]{\@gobble}%
\providecommand \bibinfo  [0]{\@secondoftwo}%
\providecommand \bibfield  [0]{\@secondoftwo}%
\providecommand \translation [1]{[#1]}%
\providecommand \BibitemOpen [0]{}%
\providecommand \bibitemStop [0]{}%
\providecommand \bibitemNoStop [0]{.\EOS\space}%
\providecommand \EOS [0]{\spacefactor3000\relax}%
\providecommand \BibitemShut  [1]{\csname bibitem#1\endcsname}%
\let\auto@bib@innerbib\@empty
\bibitem [{\citenamefont {Nagaosa}\ \emph {et~al.}(2010)\citenamefont {Nagaosa}, \citenamefont {Sinova}, \citenamefont {Onoda}, \citenamefont {MacDonald},\ and\ \citenamefont {Ong}}]{RevModPhys.82.1539}%
  \BibitemOpen
  \bibfield  {author} {\bibinfo {author} {\bibfnamefont {N.}~\bibnamefont {Nagaosa}}, \bibinfo {author} {\bibfnamefont {J.}~\bibnamefont {Sinova}}, \bibinfo {author} {\bibfnamefont {S.}~\bibnamefont {Onoda}}, \bibinfo {author} {\bibfnamefont {A.~H.}\ \bibnamefont {MacDonald}},\ and\ \bibinfo {author} {\bibfnamefont {N.~P.}\ \bibnamefont {Ong}},\ }\bibfield  {title} {\bibinfo {title} {{Anomalous} {Hall} effect},\ }\href {https://doi.org/10.1103/RevModPhys.82.1539} {\bibfield  {journal} {\bibinfo  {journal} {Rev. Mod. Phys.}\ }\textbf {\bibinfo {volume} {82}},\ \bibinfo {pages} {1539} (\bibinfo {year} {2010})}\BibitemShut {NoStop}%
\bibitem [{\citenamefont {Murakami}\ \emph {et~al.}(2003)\citenamefont {Murakami}, \citenamefont {Nagaosa},\ and\ \citenamefont {Zhang}}]{Murakami_ISH}%
  \BibitemOpen
  \bibfield  {author} {\bibinfo {author} {\bibfnamefont {S.}~\bibnamefont {Murakami}}, \bibinfo {author} {\bibfnamefont {N.}~\bibnamefont {Nagaosa}},\ and\ \bibinfo {author} {\bibfnamefont {S.-C.}\ \bibnamefont {Zhang}},\ }\bibfield  {title} {\bibinfo {title} {{Dissipationless} {Quantum} {Spin} {Current} at {Room} {Temperature}},\ }\href {https://www.science.org/doi/abs/10.1126/science.1087128} {\bibfield  {journal} {\bibinfo  {journal} {Science}\ }\textbf {\bibinfo {volume} {301}},\ \bibinfo {pages} {1348} (\bibinfo {year} {2003})}\BibitemShut {NoStop}%
\bibitem [{\citenamefont {Sinova}\ \emph {et~al.}(2015)\citenamefont {Sinova}, \citenamefont {Valenzuela}, \citenamefont {Wunderlich}, \citenamefont {Back},\ and\ \citenamefont {Jungwirth}}]{RevModPhys.87.1213}%
  \BibitemOpen
  \bibfield  {author} {\bibinfo {author} {\bibfnamefont {J.}~\bibnamefont {Sinova}}, \bibinfo {author} {\bibfnamefont {S.~O.}\ \bibnamefont {Valenzuela}}, \bibinfo {author} {\bibfnamefont {J.}~\bibnamefont {Wunderlich}}, \bibinfo {author} {\bibfnamefont {C.~H.}\ \bibnamefont {Back}},\ and\ \bibinfo {author} {\bibfnamefont {T.}~\bibnamefont {Jungwirth}},\ }\bibfield  {title} {\bibinfo {title} {{Spin} {Hall} effects},\ }\href {https://doi.org/10.1103/RevModPhys.87.1213} {\bibfield  {journal} {\bibinfo  {journal} {Rev. Mod. Phys.}\ }\textbf {\bibinfo {volume} {87}},\ \bibinfo {pages} {1213} (\bibinfo {year} {2015})}\BibitemShut {NoStop}%
\bibitem [{\citenamefont {Mizuguchi}\ and\ \citenamefont {Nakatsuji}(2019)}]{Mizuguchi31122019}%
  \BibitemOpen
  \bibfield  {author} {\bibinfo {author} {\bibfnamefont {M.}~\bibnamefont {Mizuguchi}}\ and\ \bibinfo {author} {\bibfnamefont {S.}~\bibnamefont {Nakatsuji}},\ }\bibfield  {title} {\bibinfo {title} {{Energy-harvesting} materials based on the anomalous {Nernst} effect},\ }\href {https://doi.org/10.1080/14686996.2019.1585143} {\bibfield  {journal} {\bibinfo  {journal} {Sci. Technol. Adv. Mater.}\ }\textbf {\bibinfo {volume} {20}},\ \bibinfo {pages} {262} (\bibinfo {year} {2019})}\BibitemShut {NoStop}%
\bibitem [{\citenamefont {Provost}\ and\ \citenamefont {Vallee}(1980)}]{Provost}%
  \BibitemOpen
  \bibfield  {author} {\bibinfo {author} {\bibfnamefont {J.~P.}\ \bibnamefont {Provost}}\ and\ \bibinfo {author} {\bibfnamefont {G.}~\bibnamefont {Vallee}},\ }\bibfield  {title} {\bibinfo {title} {{Riemannian} structure on manifolds of quantum states},\ }\href {https://doi.org/10.1007/BF02193559} {\bibfield  {journal} {\bibinfo  {journal} {Commun. Math. Phys.}\ }\textbf {\bibinfo {volume} {76}},\ \bibinfo {pages} {289} (\bibinfo {year} {1980})}\BibitemShut {NoStop}%
\bibitem [{\citenamefont {Xiao}\ \emph {et~al.}(2010)\citenamefont {Xiao}, \citenamefont {Chang},\ and\ \citenamefont {Niu}}]{RevModPhys.82.1959}%
  \BibitemOpen
  \bibfield  {author} {\bibinfo {author} {\bibfnamefont {D.}~\bibnamefont {Xiao}}, \bibinfo {author} {\bibfnamefont {M.-C.}\ \bibnamefont {Chang}},\ and\ \bibinfo {author} {\bibfnamefont {Q.}~\bibnamefont {Niu}},\ }\bibfield  {title} {\bibinfo {title} {{Berry} phase effects on electronic properties},\ }\href {https://doi.org/10.1103/RevModPhys.82.1959} {\bibfield  {journal} {\bibinfo  {journal} {Rev. Mod. Phys.}\ }\textbf {\bibinfo {volume} {82}},\ \bibinfo {pages} {1959} (\bibinfo {year} {2010})}\BibitemShut {NoStop}%
\bibitem [{\citenamefont {Morimoto}\ \emph {et~al.}(2023)\citenamefont {Morimoto}, \citenamefont {Kitamura},\ and\ \citenamefont {Nagaosa}}]{Morimoto_JPSJ}%
  \BibitemOpen
  \bibfield  {author} {\bibinfo {author} {\bibfnamefont {T.}~\bibnamefont {Morimoto}}, \bibinfo {author} {\bibfnamefont {S.}~\bibnamefont {Kitamura}},\ and\ \bibinfo {author} {\bibfnamefont {N.}~\bibnamefont {Nagaosa}},\ }\bibfield  {title} {\bibinfo {title} {{Geometric} {Aspects} of {Nonlinear} and {Nonequilibrium} {Phenomena}},\ }\href {https://doi.org/10.7566/JPSJ.92.072001} {\bibfield  {journal} {\bibinfo  {journal} {J. Phys. Soc. Jpn}\ }\textbf {\bibinfo {volume} {92}},\ \bibinfo {pages} {072001} (\bibinfo {year} {2023})}\BibitemShut {NoStop}%
\bibitem [{\citenamefont {Jungwirth}\ \emph {et~al.}(2002)\citenamefont {Jungwirth}, \citenamefont {Niu},\ and\ \citenamefont {MacDonald}}]{PhysRevLett.88.207208}%
  \BibitemOpen
  \bibfield  {author} {\bibinfo {author} {\bibfnamefont {T.}~\bibnamefont {Jungwirth}}, \bibinfo {author} {\bibfnamefont {Q.}~\bibnamefont {Niu}},\ and\ \bibinfo {author} {\bibfnamefont {A.~H.}\ \bibnamefont {MacDonald}},\ }\bibfield  {title} {\bibinfo {title} {{Anomalous} {Hall} {Effect} in {Ferromagnetic} {Semiconductors}},\ }\href {https://doi.org/10.1103/PhysRevLett.88.207208} {\bibfield  {journal} {\bibinfo  {journal} {Phys. Rev. Lett.}\ }\textbf {\bibinfo {volume} {88}},\ \bibinfo {pages} {207208} (\bibinfo {year} {2002})}\BibitemShut {NoStop}%
\bibitem [{\citenamefont {Onoda}\ and\ \citenamefont {Nagaosa}(2002)}]{doi:10.1143/JPSJ.71.19}%
  \BibitemOpen
  \bibfield  {author} {\bibinfo {author} {\bibfnamefont {M.}~\bibnamefont {Onoda}}\ and\ \bibinfo {author} {\bibfnamefont {N.}~\bibnamefont {Nagaosa}},\ }\bibfield  {title} {\bibinfo {title} {{Topological} {Nature} of {Anomalous} {Hall} {Effect} in {Ferromagnets}},\ }\href {https://doi.org/10.1143/JPSJ.71.19} {\bibfield  {journal} {\bibinfo  {journal} {J. Phys. Soc. Jpn.}\ }\textbf {\bibinfo {volume} {71}},\ \bibinfo {pages} {19} (\bibinfo {year} {2002})}\BibitemShut {NoStop}%
\bibitem [{\citenamefont {{\v S}mejkal}\ \emph {et~al.}(2022)\citenamefont {{\v S}mejkal}, \citenamefont {MacDonald}, \citenamefont {Sinova}, \citenamefont {Nakatsuji},\ and\ \citenamefont {Jungwirth}}]{Smejkal_AHAF}%
  \BibitemOpen
  \bibfield  {author} {\bibinfo {author} {\bibfnamefont {L.}~\bibnamefont {{\v S}mejkal}}, \bibinfo {author} {\bibfnamefont {A.~H.}\ \bibnamefont {MacDonald}}, \bibinfo {author} {\bibfnamefont {J.}~\bibnamefont {Sinova}}, \bibinfo {author} {\bibfnamefont {S.}~\bibnamefont {Nakatsuji}},\ and\ \bibinfo {author} {\bibfnamefont {T.}~\bibnamefont {Jungwirth}},\ }\bibfield  {title} {\bibinfo {title} {{Anomalous} {Hall} antiferromagnets},\ }\href {https://doi.org/10.1038/s41578-022-00430-3} {\bibfield  {journal} {\bibinfo  {journal} {Nat. Rev. Mater.}\ }\textbf {\bibinfo {volume} {7}},\ \bibinfo {pages} {482} (\bibinfo {year} {2022})}\BibitemShut {NoStop}%
\bibitem [{\citenamefont {Chen}\ \emph {et~al.}(2014)\citenamefont {Chen}, \citenamefont {Niu},\ and\ \citenamefont {MacDonald}}]{PhysRevLett.112.017205}%
  \BibitemOpen
  \bibfield  {author} {\bibinfo {author} {\bibfnamefont {H.}~\bibnamefont {Chen}}, \bibinfo {author} {\bibfnamefont {Q.}~\bibnamefont {Niu}},\ and\ \bibinfo {author} {\bibfnamefont {A.~H.}\ \bibnamefont {MacDonald}},\ }\bibfield  {title} {\bibinfo {title} {{Anomalous} {Hall} {Effect} {Arising} from {Noncollinear} {Antiferromagnetism}},\ }\href {https://doi.org/10.1103/PhysRevLett.112.017205} {\bibfield  {journal} {\bibinfo  {journal} {Phys. Rev. Lett.}\ }\textbf {\bibinfo {volume} {112}},\ \bibinfo {pages} {017205} (\bibinfo {year} {2014})}\BibitemShut {NoStop}%
\bibitem [{\citenamefont {K^^c3^^bcbler}\ and\ \citenamefont {Felser}(2014)}]{Kubler_2014}%
  \BibitemOpen
  \bibfield  {author} {\bibinfo {author} {\bibfnamefont {J.}~\bibnamefont {K^^c3^^bcbler}}\ and\ \bibinfo {author} {\bibfnamefont {C.}~\bibnamefont {Felser}},\ }\bibfield  {title} {\bibinfo {title} {{Non-collinear} antiferromagnets and the anomalous {Hall} effect},\ }\href {https://doi.org/10.1209/0295-5075/108/67001} {\bibfield  {journal} {\bibinfo  {journal} {EPL.}\ }\textbf {\bibinfo {volume} {108}},\ \bibinfo {pages} {67001} (\bibinfo {year} {2014})}\BibitemShut {NoStop}%
\bibitem [{\citenamefont {Nakatsuji}\ \emph {et~al.}(2015)\citenamefont {Nakatsuji}, \citenamefont {Kiyohara},\ and\ \citenamefont {Higo}}]{Nakatsuji_Mn3Sn}%
  \BibitemOpen
  \bibfield  {author} {\bibinfo {author} {\bibfnamefont {S.}~\bibnamefont {Nakatsuji}}, \bibinfo {author} {\bibfnamefont {N.}~\bibnamefont {Kiyohara}},\ and\ \bibinfo {author} {\bibfnamefont {T.}~\bibnamefont {Higo}},\ }\bibfield  {title} {\bibinfo {title} {{Large} anomalous {Hall} effect in a non-collinear antiferromagnet at room temperature},\ }\href {https://doi.org/10.1038/nature15723} {\bibfield  {journal} {\bibinfo  {journal} {Nature}\ }\textbf {\bibinfo {volume} {527}},\ \bibinfo {pages} {212} (\bibinfo {year} {2015})}\BibitemShut {NoStop}%
\bibitem [{\citenamefont {Kiyohara}\ \emph {et~al.}(2016)\citenamefont {Kiyohara}, \citenamefont {Tomita},\ and\ \citenamefont {Nakatsuji}}]{PhysRevApplied.5.064009}%
  \BibitemOpen
  \bibfield  {author} {\bibinfo {author} {\bibfnamefont {N.}~\bibnamefont {Kiyohara}}, \bibinfo {author} {\bibfnamefont {T.}~\bibnamefont {Tomita}},\ and\ \bibinfo {author} {\bibfnamefont {S.}~\bibnamefont {Nakatsuji}},\ }\bibfield  {title} {\bibinfo {title} {{Giant} {Anomalous} {Hall} {Effect} in the {Chiral} {Antiferromagnet} {{Mn}$_3${Ge}}},\ }\href {https://doi.org/10.1103/PhysRevApplied.5.064009} {\bibfield  {journal} {\bibinfo  {journal} {Phys. Rev. Appl.}\ }\textbf {\bibinfo {volume} {5}},\ \bibinfo {pages} {064009} (\bibinfo {year} {2016})}\BibitemShut {NoStop}%
\bibitem [{\citenamefont {Nayak}\ \emph {et~al.}(2016)\citenamefont {Nayak}, \citenamefont {Fischer}, \citenamefont {Sun}, \citenamefont {Yan}, \citenamefont {Karel}, \citenamefont {Komarek}, \citenamefont {Shekhar}, \citenamefont {Kumar}, \citenamefont {Schnelle}, \citenamefont {K{\"u}bler}, \citenamefont {Felser},\ and\ \citenamefont {Parkin}}]{doi:10.1126/sciadv.1501870}%
  \BibitemOpen
  \bibfield  {author} {\bibinfo {author} {\bibfnamefont {A.~K.}\ \bibnamefont {Nayak}}, \bibinfo {author} {\bibfnamefont {J.~E.}\ \bibnamefont {Fischer}}, \bibinfo {author} {\bibfnamefont {Y.}~\bibnamefont {Sun}}, \bibinfo {author} {\bibfnamefont {B.}~\bibnamefont {Yan}}, \bibinfo {author} {\bibfnamefont {J.}~\bibnamefont {Karel}}, \bibinfo {author} {\bibfnamefont {A.~C.}\ \bibnamefont {Komarek}}, \bibinfo {author} {\bibfnamefont {C.}~\bibnamefont {Shekhar}}, \bibinfo {author} {\bibfnamefont {N.}~\bibnamefont {Kumar}}, \bibinfo {author} {\bibfnamefont {W.}~\bibnamefont {Schnelle}}, \bibinfo {author} {\bibfnamefont {J.}~\bibnamefont {K{\"u}bler}}, \bibinfo {author} {\bibfnamefont {C.}~\bibnamefont {Felser}},\ and\ \bibinfo {author} {\bibfnamefont {S.~S.~P.}\ \bibnamefont {Parkin}},\ }\bibfield  {title} {\bibinfo {title} {{Large} anomalous {Hall} effect driven by a nonvanishing {Berry} curvature in the noncolinear antiferromagnet {Mn$_3$Ge}},\ }\href {https://doi.org/10.1126/sciadv.1501870} {\bibfield  {journal} {\bibinfo  {journal} {Science Advances}\ }\textbf {\bibinfo {volume} {2}},\ \bibinfo {pages} {e1501870} (\bibinfo {year} {2016})}\BibitemShut {NoStop}%
\bibitem [{\citenamefont {Suzuki}\ \emph {et~al.}(2017)\citenamefont {Suzuki}, \citenamefont {Koretsune}, \citenamefont {Ochi},\ and\ \citenamefont {Arita}}]{PhysRevB.95.094406}%
  \BibitemOpen
  \bibfield  {author} {\bibinfo {author} {\bibfnamefont {M.-T.}\ \bibnamefont {Suzuki}}, \bibinfo {author} {\bibfnamefont {T.}~\bibnamefont {Koretsune}}, \bibinfo {author} {\bibfnamefont {M.}~\bibnamefont {Ochi}},\ and\ \bibinfo {author} {\bibfnamefont {R.}~\bibnamefont {Arita}},\ }\bibfield  {title} {\bibinfo {title} {{Cluster} multipole theory for anomalous {Hall} effect in antiferromagnets},\ }\href {https://doi.org/10.1103/PhysRevB.95.094406} {\bibfield  {journal} {\bibinfo  {journal} {Phys. Rev. B}\ }\textbf {\bibinfo {volume} {95}},\ \bibinfo {pages} {094406} (\bibinfo {year} {2017})}\BibitemShut {NoStop}%
\bibitem [{\citenamefont {Ghimire}\ \emph {et~al.}(2018)\citenamefont {Ghimire}, \citenamefont {Botana}, \citenamefont {Jiang}, \citenamefont {Zhang}, \citenamefont {Chen},\ and\ \citenamefont {Mitchell}}]{Ghimire_AHE}%
  \BibitemOpen
  \bibfield  {author} {\bibinfo {author} {\bibfnamefont {N.~J.}\ \bibnamefont {Ghimire}}, \bibinfo {author} {\bibfnamefont {A.~S.}\ \bibnamefont {Botana}}, \bibinfo {author} {\bibfnamefont {J.~S.}\ \bibnamefont {Jiang}}, \bibinfo {author} {\bibfnamefont {J.}~\bibnamefont {Zhang}}, \bibinfo {author} {\bibfnamefont {Y.~S.}\ \bibnamefont {Chen}},\ and\ \bibinfo {author} {\bibfnamefont {J.~F.}\ \bibnamefont {Mitchell}},\ }\bibfield  {title} {\bibinfo {title} {{Large} anomalous {Hall} effect in the chiral-lattice antiferromagnet {Co}{Nb}$_3${S}$_6$},\ }\href {https://doi.org/10.1038/s41467-018-05756-7} {\bibfield  {journal} {\bibinfo  {journal} {Nat. Commun.}\ }\textbf {\bibinfo {volume} {9}},\ \bibinfo {pages} {3280} (\bibinfo {year} {2018})}\BibitemShut {NoStop}%
\bibitem [{\citenamefont {Takagi}\ \emph {et~al.}(2023)\citenamefont {Takagi}, \citenamefont {Takagi}, \citenamefont {Minami}, \citenamefont {Nomoto}, \citenamefont {Ohishi}, \citenamefont {Suzuki}, \citenamefont {Yanagi}, \citenamefont {Hirayama}, \citenamefont {Khanh}, \citenamefont {Karube}, \citenamefont {Saito}, \citenamefont {Hashizume}, \citenamefont {Kiyanagi}, \citenamefont {Tokura}, \citenamefont {Arita}, \citenamefont {Nakajima},\ and\ \citenamefont {Seki}}]{Takagi_AHE}%
  \BibitemOpen
  \bibfield  {author} {\bibinfo {author} {\bibfnamefont {H.}~\bibnamefont {Takagi}}, \bibinfo {author} {\bibfnamefont {R.}~\bibnamefont {Takagi}}, \bibinfo {author} {\bibfnamefont {S.}~\bibnamefont {Minami}}, \bibinfo {author} {\bibfnamefont {T.}~\bibnamefont {Nomoto}}, \bibinfo {author} {\bibfnamefont {K.}~\bibnamefont {Ohishi}}, \bibinfo {author} {\bibfnamefont {M.~T.}\ \bibnamefont {Suzuki}}, \bibinfo {author} {\bibfnamefont {Y.}~\bibnamefont {Yanagi}}, \bibinfo {author} {\bibfnamefont {M.}~\bibnamefont {Hirayama}}, \bibinfo {author} {\bibfnamefont {N.~D.}\ \bibnamefont {Khanh}}, \bibinfo {author} {\bibfnamefont {K.}~\bibnamefont {Karube}}, \bibinfo {author} {\bibfnamefont {H.}~\bibnamefont {Saito}}, \bibinfo {author} {\bibfnamefont {D.}~\bibnamefont {Hashizume}}, \bibinfo {author} {\bibfnamefont {R.}~\bibnamefont {Kiyanagi}}, \bibinfo {author} {\bibfnamefont {Y.}~\bibnamefont {Tokura}}, \bibinfo {author} {\bibfnamefont {R.}~\bibnamefont {Arita}}, \bibinfo {author} {\bibfnamefont {T.}~\bibnamefont {Nakajima}},\ and\ \bibinfo {author} {\bibfnamefont {S.}~\bibnamefont {Seki}},\ }\bibfield  {title} {\bibinfo {title} {{Spontaneous} topological {Hall} effect induced by non-coplanar antiferromagnetic order in intercalated van der waals materials},\ }\href {https://doi.org/10.1038/s41567-023-02017-3} {\bibfield  {journal} {\bibinfo  {journal} {Nat. Phys.}\ }\textbf {\bibinfo {volume} {19}},\ \bibinfo {pages} {961} (\bibinfo {year} {2023})}\BibitemShut {NoStop}%
\bibitem [{\citenamefont {Gonzalez~Betancourt}\ \emph {et~al.}(2023)\citenamefont {Gonzalez~Betancourt}, \citenamefont {Zub\'a{\v{c}}}, \citenamefont {{\v{Z}}elezn\'y},\ and\ \citenamefont {Kriegner}}]{PhysRevLett.130.036702}%
  \BibitemOpen
  \bibfield  {author} {\bibinfo {author} {\bibfnamefont {R.~D.}\ \bibnamefont {Gonzalez~Betancourt}}, \bibinfo {author} {\bibfnamefont {J.}~\bibnamefont {Zub\'a{\v{c}}}}, \bibinfo {author} {\bibfnamefont {J.}~\bibnamefont {{\v{Z}}elezn\'y}},\ and\ \bibinfo {author} {\bibfnamefont {D.}~\bibnamefont {Kriegner}},\ }\bibfield  {title} {\bibinfo {title} {{Spontaneous} {Anomalous} {Hall} {Effect} {Arising} from an {Unconventional} {Compensated} {Magnetic} {Phase} in a {Semiconductor}},\ }\href {https://doi.org/10.1103/PhysRevLett.130.036702} {\bibfield  {journal} {\bibinfo  {journal} {Phys. Rev. Lett.}\ }\textbf {\bibinfo {volume} {130}},\ \bibinfo {pages} {036702} (\bibinfo {year} {2023})}\BibitemShut {NoStop}%
\bibitem [{\citenamefont {Kotegawa}\ \emph {et~al.}(2023)\citenamefont {Kotegawa}, \citenamefont {Kuwata}, \citenamefont {Huyen}, \citenamefont {Arai}, \citenamefont {Tou}, \citenamefont {Matsuda}, \citenamefont {Takeda}, \citenamefont {Sugawara},\ and\ \citenamefont {Suzuki}}]{Kotegawa_NbMnP}%
  \BibitemOpen
  \bibfield  {author} {\bibinfo {author} {\bibfnamefont {H.}~\bibnamefont {Kotegawa}}, \bibinfo {author} {\bibfnamefont {Y.}~\bibnamefont {Kuwata}}, \bibinfo {author} {\bibfnamefont {V.~T.~N.}\ \bibnamefont {Huyen}}, \bibinfo {author} {\bibfnamefont {Y.}~\bibnamefont {Arai}}, \bibinfo {author} {\bibfnamefont {H.}~\bibnamefont {Tou}}, \bibinfo {author} {\bibfnamefont {M.}~\bibnamefont {Matsuda}}, \bibinfo {author} {\bibfnamefont {K.}~\bibnamefont {Takeda}}, \bibinfo {author} {\bibfnamefont {H.}~\bibnamefont {Sugawara}},\ and\ \bibinfo {author} {\bibfnamefont {M.-T.}\ \bibnamefont {Suzuki}},\ }\bibfield  {title} {\bibinfo {title} {{Large} anomalous {Hall} effect and unusual domain switching in an orthorhombic antiferromagnetic material {Nb}{Mn}{P}},\ }\href {https://doi.org/10.1038/s41535-023-00587-2} {\bibfield  {journal} {\bibinfo  {journal} {npj Quan. Mater.}\ }\textbf {\bibinfo {volume} {8}},\ \bibinfo {pages} {56} (\bibinfo {year} {2023})}\BibitemShut {NoStop}%
\bibitem [{\citenamefont {Arai}\ \emph {et~al.}(2024)\citenamefont {Arai}, \citenamefont {Hayashi}, \citenamefont {Takeda}, \citenamefont {Tou}, \citenamefont {Sugawara},\ and\ \citenamefont {Kotegawa}}]{doi:10.7566/JPSJ.93.063702}%
  \BibitemOpen
  \bibfield  {author} {\bibinfo {author} {\bibfnamefont {Y.}~\bibnamefont {Arai}}, \bibinfo {author} {\bibfnamefont {J.}~\bibnamefont {Hayashi}}, \bibinfo {author} {\bibfnamefont {K.}~\bibnamefont {Takeda}}, \bibinfo {author} {\bibfnamefont {H.}~\bibnamefont {Tou}}, \bibinfo {author} {\bibfnamefont {H.}~\bibnamefont {Sugawara}},\ and\ \bibinfo {author} {\bibfnamefont {H.}~\bibnamefont {Kotegawa}},\ }\bibfield  {title} {\bibinfo {title} {{Intrinsic} {Anomalous} {Hall} {Effect} {Arising} from {Antiferromagnetism} as {Revealed} by {High}-{Quality} {Nb}{Mn}{P}},\ }\href {https://doi.org/10.7566/JPSJ.93.063702} {\bibfield  {journal} {\bibinfo  {journal} {J. Phys. Soc. Jpn.}\ }\textbf {\bibinfo {volume} {93}},\ \bibinfo {pages} {063702} (\bibinfo {year} {2024})},\ \Eprint {https://arxiv.org/abs/https://doi.org/10.7566/JPSJ.93.063702} {https://doi.org/10.7566/JPSJ.93.063702} \BibitemShut {NoStop}%
\bibitem [{\citenamefont {Matsuda}\ \emph {et~al.}(2021)\citenamefont {Matsuda}, \citenamefont {Zhang}, \citenamefont {Kuwata}, \citenamefont {Zhang}, \citenamefont {Sakurai}, \citenamefont {Ohta}, \citenamefont {Sugawara}, \citenamefont {Takeda}, \citenamefont {Hayashi},\ and\ \citenamefont {Kotegawa}}]{PhysRevB.104.174413}%
  \BibitemOpen
  \bibfield  {author} {\bibinfo {author} {\bibfnamefont {M.}~\bibnamefont {Matsuda}}, \bibinfo {author} {\bibfnamefont {D.}~\bibnamefont {Zhang}}, \bibinfo {author} {\bibfnamefont {Y.}~\bibnamefont {Kuwata}}, \bibinfo {author} {\bibfnamefont {Q.}~\bibnamefont {Zhang}}, \bibinfo {author} {\bibfnamefont {T.}~\bibnamefont {Sakurai}}, \bibinfo {author} {\bibfnamefont {H.}~\bibnamefont {Ohta}}, \bibinfo {author} {\bibfnamefont {H.}~\bibnamefont {Sugawara}}, \bibinfo {author} {\bibfnamefont {K.}~\bibnamefont {Takeda}}, \bibinfo {author} {\bibfnamefont {J.}~\bibnamefont {Hayashi}},\ and\ \bibinfo {author} {\bibfnamefont {H.}~\bibnamefont {Kotegawa}},\ }\bibfield  {title} {\bibinfo {title} {{Noncollinear} spin structure with weak ferromagnetism in {NbMnP}},\ }\href {https://link.aps.org/doi/10.1103/PhysRevB.104.174413} {\bibfield  {journal} {\bibinfo  {journal} {Phys. Rev. B}\ }\textbf {\bibinfo {volume} {104}},\ \bibinfo {pages} {174413} (\bibinfo {year} {2021})}\BibitemShut {NoStop}%
\bibitem [{\citenamefont {Momma}\ and\ \citenamefont {Izumi}(2011)}]{Momma:db5098}%
  \BibitemOpen
  \bibfield  {author} {\bibinfo {author} {\bibfnamefont {K.}~\bibnamefont {Momma}}\ and\ \bibinfo {author} {\bibfnamefont {F.}~\bibnamefont {Izumi}},\ }\bibfield  {title} {\bibinfo {title} {{{\it VESTA3} for three-dimensional visualization of crystal, volumetric and morphology data}},\ }\href {https://doi.org/10.1107/S0021889811038970} {\bibfield  {journal} {\bibinfo  {journal} {J. Appl. Crystallogr.}\ }\textbf {\bibinfo {volume} {44}},\ \bibinfo {pages} {1272} (\bibinfo {year} {2011})}\BibitemShut {NoStop}%
\bibitem [{\citenamefont {Wang}\ \emph {et~al.}(2021)\citenamefont {Wang}, \citenamefont {Gao},\ and\ \citenamefont {Xiao}}]{PhysRevLett.127.277201}%
  \BibitemOpen
  \bibfield  {author} {\bibinfo {author} {\bibfnamefont {C.}~\bibnamefont {Wang}}, \bibinfo {author} {\bibfnamefont {Y.}~\bibnamefont {Gao}},\ and\ \bibinfo {author} {\bibfnamefont {D.}~\bibnamefont {Xiao}},\ }\bibfield  {title} {\bibinfo {title} {{Intrinsic} {Nonlinear} {Hall} {Effect} in {Antiferromagnetic} {Tetragonal} {CuMnAs}},\ }\href {https://doi.org/10.1103/PhysRevLett.127.277201} {\bibfield  {journal} {\bibinfo  {journal} {Phys. Rev. Lett.}\ }\textbf {\bibinfo {volume} {127}},\ \bibinfo {pages} {277201} (\bibinfo {year} {2021})}\BibitemShut {NoStop}%
\bibitem [{\citenamefont {Liu}\ \emph {et~al.}(2021)\citenamefont {Liu}, \citenamefont {Zhao}, \citenamefont {Huang}, \citenamefont {Wu}, \citenamefont {Sheng}, \citenamefont {Xiao},\ and\ \citenamefont {Yang}}]{PhysRevLett.127.277202}%
  \BibitemOpen
  \bibfield  {author} {\bibinfo {author} {\bibfnamefont {H.}~\bibnamefont {Liu}}, \bibinfo {author} {\bibfnamefont {J.}~\bibnamefont {Zhao}}, \bibinfo {author} {\bibfnamefont {Y.-X.}\ \bibnamefont {Huang}}, \bibinfo {author} {\bibfnamefont {W.}~\bibnamefont {Wu}}, \bibinfo {author} {\bibfnamefont {X.-L.}\ \bibnamefont {Sheng}}, \bibinfo {author} {\bibfnamefont {C.}~\bibnamefont {Xiao}},\ and\ \bibinfo {author} {\bibfnamefont {S.~A.}\ \bibnamefont {Yang}},\ }\bibfield  {title} {\bibinfo {title} {{Intrinsic} {Second}-{Order} {Anomalous} {Hall} {Effect} and {Its} {Application} in {Compensated} {Antiferromagnets}},\ }\href {https://doi.org/10.1103/PhysRevLett.127.277202} {\bibfield  {journal} {\bibinfo  {journal} {Phys. Rev. Lett.}\ }\textbf {\bibinfo {volume} {127}},\ \bibinfo {pages} {277202} (\bibinfo {year} {2021})}\BibitemShut {NoStop}%
\bibitem [{\citenamefont {Gao}\ \emph {et~al.}(2023)\citenamefont {Gao}, \citenamefont {Liu}, \citenamefont {Qiu}, \citenamefont {Ghosh}, \citenamefont {Trevisan}, \citenamefont {Onishi}, \citenamefont {Hu}, \citenamefont {Qian}, \citenamefont {Tien}, \citenamefont {Chen}, \citenamefont {Huang}, \citenamefont {B{\'e}rub{\'e}}, \citenamefont {Li}, \citenamefont {Tzschaschel}, \citenamefont {Dinh}, \citenamefont {Sun}, \citenamefont {Ho}, \citenamefont {Lien}, \citenamefont {Singh}, \citenamefont {Watanabe}, \citenamefont {Taniguchi}, \citenamefont {Bell}, \citenamefont {Lin}, \citenamefont {Chang}, \citenamefont {Du}, \citenamefont {Bansil}, \citenamefont {Fu}, \citenamefont {Ni}, \citenamefont {Orth}, \citenamefont {Ma},\ and\ \citenamefont {Xu}}]{doi:10.1126/science.adf1506}%
  \BibitemOpen
  \bibfield  {author} {\bibinfo {author} {\bibfnamefont {A.}~\bibnamefont {Gao}}, \bibinfo {author} {\bibfnamefont {Y.-F.}\ \bibnamefont {Liu}}, \bibinfo {author} {\bibfnamefont {J.-X.}\ \bibnamefont {Qiu}}, \bibinfo {author} {\bibfnamefont {B.}~\bibnamefont {Ghosh}}, \bibinfo {author} {\bibfnamefont {T.~V.}\ \bibnamefont {Trevisan}}, \bibinfo {author} {\bibfnamefont {Y.}~\bibnamefont {Onishi}}, \bibinfo {author} {\bibfnamefont {C.}~\bibnamefont {Hu}}, \bibinfo {author} {\bibfnamefont {T.}~\bibnamefont {Qian}}, \bibinfo {author} {\bibfnamefont {H.-J.}\ \bibnamefont {Tien}}, \bibinfo {author} {\bibfnamefont {S.-W.}\ \bibnamefont {Chen}}, \bibinfo {author} {\bibfnamefont {M.}~\bibnamefont {Huang}}, \bibinfo {author} {\bibfnamefont {D.}~\bibnamefont {B{\'e}rub{\'e}}}, \bibinfo {author} {\bibfnamefont {H.}~\bibnamefont {Li}}, \bibinfo {author} {\bibfnamefont {C.}~\bibnamefont {Tzschaschel}}, \bibinfo {author} {\bibfnamefont {T.}~\bibnamefont {Dinh}}, \bibinfo {author} {\bibfnamefont {Z.}~\bibnamefont {Sun}}, \bibinfo {author} {\bibfnamefont {S.-C.}\ \bibnamefont {Ho}}, \bibinfo {author} {\bibfnamefont {S.-W.}\ \bibnamefont {Lien}}, \bibinfo {author} {\bibfnamefont {B.}~\bibnamefont {Singh}}, \bibinfo {author} {\bibfnamefont {K.}~\bibnamefont {Watanabe}}, \bibinfo {author} {\bibfnamefont {T.}~\bibnamefont {Taniguchi}}, \bibinfo {author} {\bibfnamefont {D.~C.}\ \bibnamefont {Bell}}, \bibinfo {author} {\bibfnamefont {H.}~\bibnamefont {Lin}}, \bibinfo {author} {\bibfnamefont {T.-R.}\ \bibnamefont {Chang}}, \bibinfo {author} {\bibfnamefont {C.~R.}\ \bibnamefont {Du}}, \bibinfo {author} {\bibfnamefont {A.}~\bibnamefont {Bansil}}, \bibinfo {author} {\bibfnamefont {L.}~\bibnamefont {Fu}}, \bibinfo {author} {\bibfnamefont {N.}~\bibnamefont {Ni}}, \bibinfo {author} {\bibfnamefont {P.~P.}\ \bibnamefont {Orth}}, \bibinfo {author} {\bibfnamefont {Q.}~\bibnamefont {Ma}},\ and\ \bibinfo {author} {\bibfnamefont {S.-Y.}\ \bibnamefont {Xu}},\ }\bibfield  {title} {\bibinfo {title} {{Quantum} metric nonlinear hall effect in a topological antiferromagnetic heterostructure},\ }\href {https://www.science.org/doi/abs/10.1126/science.adf1506} {\bibfield  {journal} {\bibinfo  {journal} {Science}\ }\textbf {\bibinfo {volume} {381}},\ \bibinfo {pages} {181} (\bibinfo {year} {2023})}\BibitemShut {NoStop}%
\bibitem [{\citenamefont {Wang}\ \emph {et~al.}(2023)\citenamefont {Wang}, \citenamefont {Kaplan}, \citenamefont {Zhang}, \citenamefont {Holder}, \citenamefont {Cao}, \citenamefont {Wang}, \citenamefont {Zhou}, \citenamefont {Zhou}, \citenamefont {Jiang}, \citenamefont {Zhang}, \citenamefont {Ru}, \citenamefont {Cai}, \citenamefont {Watanabe}, \citenamefont {Taniguchi}, \citenamefont {Yan},\ and\ \citenamefont {Gao}}]{Wang_Nature}%
  \BibitemOpen
  \bibfield  {author} {\bibinfo {author} {\bibfnamefont {N.}~\bibnamefont {Wang}}, \bibinfo {author} {\bibfnamefont {D.}~\bibnamefont {Kaplan}}, \bibinfo {author} {\bibfnamefont {Z.}~\bibnamefont {Zhang}}, \bibinfo {author} {\bibfnamefont {T.}~\bibnamefont {Holder}}, \bibinfo {author} {\bibfnamefont {N.}~\bibnamefont {Cao}}, \bibinfo {author} {\bibfnamefont {A.}~\bibnamefont {Wang}}, \bibinfo {author} {\bibfnamefont {X.}~\bibnamefont {Zhou}}, \bibinfo {author} {\bibfnamefont {F.}~\bibnamefont {Zhou}}, \bibinfo {author} {\bibfnamefont {Z.}~\bibnamefont {Jiang}}, \bibinfo {author} {\bibfnamefont {C.}~\bibnamefont {Zhang}}, \bibinfo {author} {\bibfnamefont {S.}~\bibnamefont {Ru}}, \bibinfo {author} {\bibfnamefont {H.}~\bibnamefont {Cai}}, \bibinfo {author} {\bibfnamefont {K.}~\bibnamefont {Watanabe}}, \bibinfo {author} {\bibfnamefont {T.}~\bibnamefont {Taniguchi}}, \bibinfo {author} {\bibfnamefont {B.}~\bibnamefont {Yan}},\ and\ \bibinfo {author} {\bibfnamefont {W.}~\bibnamefont {Gao}},\ }\bibfield  {title} {\bibinfo {title} {{Quantum}-metric-induced nonlinear transport in a topological antiferromagnet},\ }\href {https://doi.org/10.1038/s41586-023-06363-3} {\bibfield  {journal} {\bibinfo  {journal} {Nature}\ }\textbf {\bibinfo {volume} {621}},\ \bibinfo {pages} {487} (\bibinfo {year} {2023})}\BibitemShut {NoStop}%
\bibitem [{\citenamefont {Das}\ \emph {et~al.}(2023)\citenamefont {Das}, \citenamefont {Lahiri}, \citenamefont {Atencia}, \citenamefont {Culcer},\ and\ \citenamefont {Agarwal}}]{PhysRevB.108.L201405}%
  \BibitemOpen
  \bibfield  {author} {\bibinfo {author} {\bibfnamefont {K.}~\bibnamefont {Das}}, \bibinfo {author} {\bibfnamefont {S.}~\bibnamefont {Lahiri}}, \bibinfo {author} {\bibfnamefont {R.~B.}\ \bibnamefont {Atencia}}, \bibinfo {author} {\bibfnamefont {D.}~\bibnamefont {Culcer}},\ and\ \bibinfo {author} {\bibfnamefont {A.}~\bibnamefont {Agarwal}},\ }\bibfield  {title} {\bibinfo {title} {Intrinsic nonlinear conductivities induced by the quantum metric},\ }\href {https://doi.org/10.1103/PhysRevB.108.L201405} {\bibfield  {journal} {\bibinfo  {journal} {Phys. Rev. B}\ }\textbf {\bibinfo {volume} {108}},\ \bibinfo {pages} {L201405} (\bibinfo {year} {2023})}\BibitemShut {NoStop}%
\bibitem [{\citenamefont {Gao}\ \emph {et~al.}(2014)\citenamefont {Gao}, \citenamefont {Yang},\ and\ \citenamefont {Niu}}]{PhysRevLett.112.166601}%
  \BibitemOpen
  \bibfield  {author} {\bibinfo {author} {\bibfnamefont {Y.}~\bibnamefont {Gao}}, \bibinfo {author} {\bibfnamefont {S.~A.}\ \bibnamefont {Yang}},\ and\ \bibinfo {author} {\bibfnamefont {Q.}~\bibnamefont {Niu}},\ }\bibfield  {title} {\bibinfo {title} {{Field} {Induced} {Positional} {Shift} of {Bloch} {Electrons} and {Its} {Dynamical} {Implications}},\ }\href {https://doi.org/10.1103/PhysRevLett.112.166601} {\bibfield  {journal} {\bibinfo  {journal} {Phys. Rev. Lett.}\ }\textbf {\bibinfo {volume} {112}},\ \bibinfo {pages} {166601} (\bibinfo {year} {2014})}\BibitemShut {NoStop}%
\bibitem [{\citenamefont {Sodemann}\ and\ \citenamefont {Fu}(2015)}]{PhysRevLett.115.216806}%
  \BibitemOpen
  \bibfield  {author} {\bibinfo {author} {\bibfnamefont {I.}~\bibnamefont {Sodemann}}\ and\ \bibinfo {author} {\bibfnamefont {L.}~\bibnamefont {Fu}},\ }\bibfield  {title} {\bibinfo {title} {{Quantum} {Nonlinear} {Hall} {Effect} {Induced} by {Berry} {Curvature} {Dipole} in {Time}-{Reversal} {Invariant} {Materials}},\ }\href {https://doi.org/10.1103/PhysRevLett.115.216806} {\bibfield  {journal} {\bibinfo  {journal} {Phys. Rev. Lett.}\ }\textbf {\bibinfo {volume} {115}},\ \bibinfo {pages} {216806} (\bibinfo {year} {2015})}\BibitemShut {NoStop}%
\bibitem [{\citenamefont {Ma}\ \emph {et~al.}(2019)\citenamefont {Ma}, \citenamefont {Xu}, \citenamefont {Shen}, \citenamefont {MacNeill}, \citenamefont {Fatemi}, \citenamefont {Chang}, \citenamefont {Mier~Valdivia}, \citenamefont {Wu}, \citenamefont {Du}, \citenamefont {Hsu}, \citenamefont {Fang}, \citenamefont {Gibson}, \citenamefont {Watanabe}, \citenamefont {Taniguchi}, \citenamefont {Cava}, \citenamefont {Kaxiras}, \citenamefont {Lu}, \citenamefont {Lin}, \citenamefont {Fu}, \citenamefont {Gedik},\ and\ \citenamefont {Jarillo-Herrero}}]{Ma_Nature7739}%
  \BibitemOpen
  \bibfield  {author} {\bibinfo {author} {\bibfnamefont {Q.}~\bibnamefont {Ma}}, \bibinfo {author} {\bibfnamefont {S.-Y.}\ \bibnamefont {Xu}}, \bibinfo {author} {\bibfnamefont {H.}~\bibnamefont {Shen}}, \bibinfo {author} {\bibfnamefont {D.}~\bibnamefont {MacNeill}}, \bibinfo {author} {\bibfnamefont {V.}~\bibnamefont {Fatemi}}, \bibinfo {author} {\bibfnamefont {T.-R.}\ \bibnamefont {Chang}}, \bibinfo {author} {\bibfnamefont {A.~M.}\ \bibnamefont {Mier~Valdivia}}, \bibinfo {author} {\bibfnamefont {S.}~\bibnamefont {Wu}}, \bibinfo {author} {\bibfnamefont {Z.}~\bibnamefont {Du}}, \bibinfo {author} {\bibfnamefont {C.-H.}\ \bibnamefont {Hsu}}, \bibinfo {author} {\bibfnamefont {S.}~\bibnamefont {Fang}}, \bibinfo {author} {\bibfnamefont {Q.~D.}\ \bibnamefont {Gibson}}, \bibinfo {author} {\bibfnamefont {K.}~\bibnamefont {Watanabe}}, \bibinfo {author} {\bibfnamefont {T.}~\bibnamefont {Taniguchi}}, \bibinfo {author} {\bibfnamefont {R.~J.}\ \bibnamefont {Cava}}, \bibinfo {author} {\bibfnamefont {E.}~\bibnamefont {Kaxiras}}, \bibinfo {author} {\bibfnamefont {H.-Z.}\ \bibnamefont {Lu}}, \bibinfo {author} {\bibfnamefont {H.}~\bibnamefont {Lin}}, \bibinfo {author} {\bibfnamefont {L.}~\bibnamefont {Fu}}, \bibinfo {author} {\bibfnamefont {N.}~\bibnamefont {Gedik}},\ and\ \bibinfo {author} {\bibfnamefont {P.}~\bibnamefont {Jarillo-Herrero}},\ }\bibfield  {title} {\bibinfo {title} {{Observation} of the nonlinear {Hall} effect under time-reversal-symmetric conditions},\ }\href {https://doi.org/10.1038/s41586-018-0807-6} {\bibfield  {journal} {\bibinfo  {journal} {Nature}\ }\textbf {\bibinfo {volume} {565}},\ \bibinfo {pages} {337} (\bibinfo {year} {2019})}\BibitemShut {NoStop}%
\bibitem [{\citenamefont {Kang}\ \emph {et~al.}(2019)\citenamefont {Kang}, \citenamefont {Li}, \citenamefont {Sohn}, \citenamefont {Shan},\ and\ \citenamefont {Mak}}]{KangNatMater4}%
  \BibitemOpen
  \bibfield  {author} {\bibinfo {author} {\bibfnamefont {K.}~\bibnamefont {Kang}}, \bibinfo {author} {\bibfnamefont {T.}~\bibnamefont {Li}}, \bibinfo {author} {\bibfnamefont {E.}~\bibnamefont {Sohn}}, \bibinfo {author} {\bibfnamefont {J.}~\bibnamefont {Shan}},\ and\ \bibinfo {author} {\bibfnamefont {K.~F.}\ \bibnamefont {Mak}},\ }\bibfield  {title} {\bibinfo {title} {{Nonlinear} anomalous {Hall} effect in few-layer {WTe$_2$}},\ }\href {https://doi.org/10.1038/s41563-019-0294-7} {\bibfield  {journal} {\bibinfo  {journal} {Nat. Mater.}\ }\textbf {\bibinfo {volume} {18}},\ \bibinfo {pages} {324} (\bibinfo {year} {2019})}\BibitemShut {NoStop}%
\bibitem [{\citenamefont {Xiao}\ \emph {et~al.}(2019)\citenamefont {Xiao}, \citenamefont {Du},\ and\ \citenamefont {Niu}}]{PhysRevB.100.165422}%
  \BibitemOpen
  \bibfield  {author} {\bibinfo {author} {\bibfnamefont {C.}~\bibnamefont {Xiao}}, \bibinfo {author} {\bibfnamefont {Z.~Z.}\ \bibnamefont {Du}},\ and\ \bibinfo {author} {\bibfnamefont {Q.}~\bibnamefont {Niu}},\ }\bibfield  {title} {\bibinfo {title} {{Theory} of nonlinear {Hall} effects: {Modified} semiclassics from quantum kinetics},\ }\href {https://doi.org/10.1103/PhysRevB.100.165422} {\bibfield  {journal} {\bibinfo  {journal} {Phys. Rev. B}\ }\textbf {\bibinfo {volume} {100}},\ \bibinfo {pages} {165422} (\bibinfo {year} {2019})}\BibitemShut {NoStop}%
\bibitem [{\citenamefont {Du}\ \emph {et~al.}(2021)\citenamefont {Du}, \citenamefont {Wang}, \citenamefont {Sun}, \citenamefont {Lu},\ and\ \citenamefont {Xie}}]{Du_NatCommun1}%
  \BibitemOpen
  \bibfield  {author} {\bibinfo {author} {\bibfnamefont {Z.~Z.}\ \bibnamefont {Du}}, \bibinfo {author} {\bibfnamefont {C.~M.}\ \bibnamefont {Wang}}, \bibinfo {author} {\bibfnamefont {H.-P.}\ \bibnamefont {Sun}}, \bibinfo {author} {\bibfnamefont {H.-Z.}\ \bibnamefont {Lu}},\ and\ \bibinfo {author} {\bibfnamefont {X.~C.}\ \bibnamefont {Xie}},\ }\bibfield  {title} {\bibinfo {title} {{Quantum} theory of the nonlinear {Hall} effect},\ }\href {https://doi.org/10.1038/s41467-021-25273-4} {\bibfield  {journal} {\bibinfo  {journal} {Nat. Commun.}\ }\textbf {\bibinfo {volume} {12}},\ \bibinfo {pages} {5038} (\bibinfo {year} {2021})}\BibitemShut {NoStop}%
\bibitem [{\citenamefont {Giannozzi}\ and\ \citenamefont {et~al}(2009)}]{Giannozzi_2009}%
  \BibitemOpen
  \bibfield  {author} {\bibinfo {author} {\bibfnamefont {P.}~\bibnamefont {Giannozzi}}\ and\ \bibinfo {author} {\bibnamefont {et~al}},\ }\bibfield  {title} {\bibinfo {title} {Quantum espresso: a modular and open-source software project for quantum simulations of materials},\ }\href {https://doi.org/10.1088/0953-8984/21/39/395502} {\bibfield  {journal} {\bibinfo  {journal} {J. Phys. Condens. Matter}\ }\textbf {\bibinfo {volume} {21}},\ \bibinfo {pages} {395502} (\bibinfo {year} {2009})}\BibitemShut {NoStop}%
\bibitem [{\citenamefont {Perdew}\ \emph {et~al.}(1996)\citenamefont {Perdew}, \citenamefont {Burke},\ and\ \citenamefont {Ernzerhof}}]{PhysRevLett.77.3865}%
  \BibitemOpen
  \bibfield  {author} {\bibinfo {author} {\bibfnamefont {J.~P.}\ \bibnamefont {Perdew}}, \bibinfo {author} {\bibfnamefont {K.}~\bibnamefont {Burke}},\ and\ \bibinfo {author} {\bibfnamefont {M.}~\bibnamefont {Ernzerhof}},\ }\bibfield  {title} {\bibinfo {title} {{Generalized} {Gradient} {Approximation} {Made} {Simple}},\ }\href {https://doi.org/10.1103/PhysRevLett.77.3865} {\bibfield  {journal} {\bibinfo  {journal} {Phys. Rev. Lett.}\ }\textbf {\bibinfo {volume} {77}},\ \bibinfo {pages} {3865} (\bibinfo {year} {1996})}\BibitemShut {NoStop}%
\bibitem [{\citenamefont {Bl\"ochl}(1994)}]{PhysRevB.50.17953}%
  \BibitemOpen
  \bibfield  {author} {\bibinfo {author} {\bibfnamefont {P.~E.}\ \bibnamefont {Bl\"ochl}},\ }\bibfield  {title} {\bibinfo {title} {{Projector} augmented-wave method},\ }\href {https://doi.org/10.1103/PhysRevB.50.17953} {\bibfield  {journal} {\bibinfo  {journal} {Phys. Rev. B}\ }\textbf {\bibinfo {volume} {50}},\ \bibinfo {pages} {17953} (\bibinfo {year} {1994})}\BibitemShut {NoStop}%
\bibitem [{\citenamefont {Kresse}\ and\ \citenamefont {Joubert}(1999)}]{PhysRevB.59.1758}%
  \BibitemOpen
  \bibfield  {author} {\bibinfo {author} {\bibfnamefont {G.}~\bibnamefont {Kresse}}\ and\ \bibinfo {author} {\bibfnamefont {D.}~\bibnamefont {Joubert}},\ }\bibfield  {title} {\bibinfo {title} {{From} ultrasoft pseudopotentials to the projector augmented-wave method},\ }\href {https://doi.org/10.1103/PhysRevB.59.1758} {\bibfield  {journal} {\bibinfo  {journal} {Phys. Rev. B}\ }\textbf {\bibinfo {volume} {59}},\ \bibinfo {pages} {1758} (\bibinfo {year} {1999})}\BibitemShut {NoStop}%
\bibitem [{\citenamefont {Corso}(2014)}]{DALCORSO2014337}%
  \BibitemOpen
  \bibfield  {author} {\bibinfo {author} {\bibfnamefont {A.~D.}\ \bibnamefont {Corso}},\ }\bibfield  {title} {\bibinfo {title} {{Pseudopotentials} periodic table: {From} {H} to {P}u},\ }\href {https://www.sciencedirect.com/science/article/pii/S0927025614005187} {\bibfield  {journal} {\bibinfo  {journal} {Comput. Mater. Science}\ }\textbf {\bibinfo {volume} {95}},\ \bibinfo {pages} {337} (\bibinfo {year} {2014})}\BibitemShut {NoStop}%
\bibitem [{\citenamefont {Pizzi}\ and\ \citenamefont {et~al}(2020)}]{Pizzi_2020}%
  \BibitemOpen
  \bibfield  {author} {\bibinfo {author} {\bibfnamefont {G.}~\bibnamefont {Pizzi}}\ and\ \bibinfo {author} {\bibnamefont {et~al}},\ }\bibfield  {title} {\bibinfo {title} {Wannier90 as a community code: new features and applications},\ }\href {https://doi.org/10.1088/1361-648X/ab51ff} {\bibfield  {journal} {\bibinfo  {journal} {J. Phys. Condens. Matter}\ }\textbf {\bibinfo {volume} {32}},\ \bibinfo {pages} {165902} (\bibinfo {year} {2020})}\BibitemShut {NoStop}%
\bibitem [{\citenamefont {Vanderbilt}(2018)}]{Vanderbilt_book_2018}%
  \BibitemOpen
  \bibfield  {author} {\bibinfo {author} {\bibfnamefont {D.}~\bibnamefont {Vanderbilt}},\ }\href@noop {} {\emph {\bibinfo {title} {{Berry} {Phases} in {Electronic} {Structure} {Theory}: {Electric} {Polarization}, {Orbital} {Magnetization} and {Topological} {Insulators}}}}\ (\bibinfo  {publisher} {Cambridge University Press},\ \bibinfo {address} {Cambridge},\ \bibinfo {year} {2018})\BibitemShut {NoStop}%
\bibitem [{\citenamefont {Huyen}\ \emph {et~al.}(2019)\citenamefont {Huyen}, \citenamefont {Suzuki}, \citenamefont {Yamauchi},\ and\ \citenamefont {Oguchi}}]{PhysRevB.100.094426}%
  \BibitemOpen
  \bibfield  {author} {\bibinfo {author} {\bibfnamefont {V.~T.~N.}\ \bibnamefont {Huyen}}, \bibinfo {author} {\bibfnamefont {M.-T.}\ \bibnamefont {Suzuki}}, \bibinfo {author} {\bibfnamefont {K.}~\bibnamefont {Yamauchi}},\ and\ \bibinfo {author} {\bibfnamefont {T.}~\bibnamefont {Oguchi}},\ }\bibfield  {title} {\bibinfo {title} {{Topology} analysis for anomalous {Hall} effect in the noncollinear antiferromagnetic states of $\mathrm{M}\mathrm{n}_{3}{A}\mathrm{N}$ $({A}=\mathrm{Ni}, \mathrm{Cu}, \mathrm{Zn}, \mathrm{Ga}, \mathrm{Ge}, \mathrm{Pd}, \mathrm{In}, \mathrm{Sn}, \mathrm{Ir}, \mathrm{Pt})$},\ }\href {https://doi.org/10.1103/PhysRevB.100.094426} {\bibfield  {journal} {\bibinfo  {journal} {Phys. Rev. B}\ }\textbf {\bibinfo {volume} {100}},\ \bibinfo {pages} {094426} (\bibinfo {year} {2019})}\BibitemShut {NoStop}%
\end{thebibliography}%

\end{document}